\documentclass[prd,aps]{revtex4}

\usepackage{amsmath}
\usepackage{graphicx}
\usepackage{amssymb}
\usepackage{color}

\newcommand{\be}{\begin{eqnarray}}
\newcommand{\ee}{\end{eqnarray}}
\newcommand{\bi}{\begin{itemize}}
\newcommand{\ei}{\end{itemize}}

\newcommand{\bx}{{\vec{x}}}

\newcounter{hran}

\def\MSbar{\relax\ifmmode\overline{\rm MS}\else{$\overline{\rm MS}${ }}\fi}
\def\del{\partial}

\def\H{{\cal H}}

\def\d{{\rm d}}

\def\h{{\rm h}}

\def\vq{\vec{q}}
\def\vx{\vec{x}}
\def\vv{\vec{v}}
\def\1{{(1)}}
\def\2{{(2)}}
\def\3{{(3)}}

\def\bx{{\vec{x}}}
\def\vk{{\vec{k}}}

\def\bnabla{{\vec{\nabla}}}
\def\bv{{\vec{v}}}
\def\t{\tau}

\def\t{\tau}

 \def\vx{\vec{ x}}
   
\def\vk{\vec{k}}

\def\vq{\vec{q}}

\def\ar{(\vx,\t)}
\def\qvh{\mathrm{\hat{\bf{q}}}}

\numberwithin{equation}{section}
\begin{document}

\def\thefootnote{\fnsymbol{footnote}}










\renewcommand{\topfraction}{0.99}
\renewcommand{\bottomfraction}{0.99}

\title{ \Large\bf Non-local halo bias with and without massive neutrinos}

\vskip 2cm

\author{Matteo Biagetti$^{(1)}$, Vincent Desjacques$^{(1)}$, Alex Kehagias$^{(1,2)}$ and Antonio Riotto$^{(1)}$}

\vskip 5cm
\address{$^{(1)}$Universit\'e de Gen\`eve, Department of Theoretical Physics and Center for Astroparticle Physics (CAP),\\ 24 quai E. Ansermet, CH-1211 Geneva 4, Switzerland}

\address{$^{(2)}$Physics Division, National Technical University of Athens, 15780 Zografou Campus, 
Athens, Greece}

\date{\today}

\begin{abstract}
\noindent
Understanding the biasing between the clustering properties of halos  and the underlying dark matter distribution 
is important for extracting cosmological information from ongoing and upcoming galaxy surveys. 
While on sufficiently larges scales the halo overdensity is a local function of the mass density fluctuations, on 
smaller scales the gravitational evolution generates non-local terms in the halo density field. 
We characterize the magnitude of these contributions at third-order in perturbation theory by identifying the 
coefficients of the non-local invariant operators, and extend our calculation to include non-local (Lagrangian) 
terms induced by a peak constraint.
We apply our results to describe the scale-dependence of halo bias in cosmologies with massive neutrinos. 
The inclusion of gravity-induced non-local terms and, especially, a Lagrangian $k^2$-contribution is essential to 
reproduce the numerical data accurately. 
We use the peak-background split to derive the numerical values of the various bias coefficients from the excursion
set peak mass function. For neutrino masses in the range $0\leq \sum_i m_{\nu_i} \leq 0.6$ eV, we are able to fit 
the data with a precision of a few percents up to $k=0.3\, h {\rm  \,Mpc^{-1}}$ without any free parameter.
\end{abstract}

\def\thefootnote{\arabic{footnote}}
\setcounter{footnote}{0}
\maketitle



\section{Introduction}\pagenumbering{arabic}
\noindent
Cosmological parameter estimation from galaxy clustering data is hampered by galaxy biasing, 
{\it i.e.} by the fact that galaxies do not perfectly trace the underlying mass distribution.
Various theoretical arguments and outcomes of numerical simulations both suggest that, on 
sufficiently large scales, the galaxy overdensity $\delta_{\rm h}(\vx,\t)$ can be written as 
a generic function $f[\delta(\vx,\t)]$ of the mass density perturbation $\delta(\vx,\t)$. 
This function can be  Taylor-expanded, with the unknown coefficients in the series defining 
the so-called  bias parameters
 \be
 \label{local}
 \delta_{\rm h}(\vx,\t)= f[\delta(\vx,\t)]
 =b_1(\t)\delta(\vx,\t)+\frac{b_2(\t)}{2}
 \left(\delta^2(\vx,\t)-\langle\delta^2(\vx,\t)\rangle\right)+\cdots.
 \ee
We use the subscript (${\rm h}$), which stands for halos, because galaxies form within Dark 
Matter (DM) halos and, therefore, understanding the clustering properties of the halos is a 
key step towards an accurate description of galaxy biasing. 
Furthermore, this is simpler problem since DM halos collapse under the action of gravity solely.
 
The local model \cite{local} described in, e.g., (\ref{local}) is however incomplete: there is 
indeed no a priori reason why the halo density contrast should be only a local function of the 
matter density contrast. 
Indeed, already at second-order in perturbation theory, the gravitational evolution generates 
a term quadratic in the tidal tensor and, therefore, non-local in the density field. This term 
is absent in the initial conditions. This point was made in Refs. \cite{peebles80,fry84,goroff86, 
 bouchet92} for the matter density contrast and subsequently investigated in the context of halo 
bias in Refs. \cite{catelan98,catelan2000,MD,des10,mat,nonlocalbias,baldauf,norena}. 
In Refs. \cite{MD,norena} in particular, it was pointed out that the symmetries (essentially 
extended Galilean and Lifshitz symmetries) present in the dynamical equations for the halo and 
DM systems allow to construct a set of invariant operators which should appear in the halo bias 
expansion, precisely because they are allowed by the symmetries of the problem. 
These invariants lead to non-local bias contributions, which numerical simulations have already 
detected at the quadratic level and found to agree well with the prediction of perturbation theory 
\cite{baldauf,nonlocalbias}. 
In this paper we compute the non-local bias coefficients in the basis of invariant non-local bias 
operators at third-order in perturbation theory by using the Lagrangian Bias parameters generated 
within the peak-background split model. 
 
Although the magnitude of the non-local bias  terms is small relative to the linear halo bias, 
upcoming large scale structure data will be sensitive to them.
In particular, the impact of the non-local bias, if not accounted for, could mimic the $k$-dependent
suppression of the growth rate  in cosmologies with massive neutrinos 
\cite{saito11,zhao13,audren13,font13,beutler13,loverde14a} (see \cite{Lesgourgues:2006nd,Lesgourgues:2012uu}
for detailed reviews of the subject). 
Their influence  on the spatial distribution of DM  halos has been recently scrutinized in a series 
of papers \cite{Villaescusa-Navarro:2013pva,Castorina:2013wga,Costanzi:2013bha}  using large N-body 
simulations that incorporate massive neutrinos as an extra set of particles (see also the recent work of \cite{loverde14b}). 
Massive neutrinos generate a scale-dependent bias in the power spectrum of DM halos. As we will see, 
this effect is somewhat degenerate with the signature left by the various non-local bias terms, which 
must be taken into account in order to reproduce the N-body data with good accuracy ({\it i.e.} at 
the $\leq$5\% level). 

The paper is organized as follows. In Sec.~\ref{sec:nonlocal}, we present our computation of the 
non-local bias at third-order in perturbation theory as well as the calculation of the Lagrangian 
bias parameters through the peak-background split model.
In Sec.~\ref{sec:neutrinobias} we discuss the halo bias in the presence of massive neutrinos and
argue that contributions from nonlocal Lagrangian and gravity bias must be accounted for in order
to fit the numerical data. We conclude in Sec.~\ref{sec:conclusion}.

 \section{Non-local bias up to third-order}\label{sec:nonlocal}
\noindent
In this section we extend the analysis of \cite{MD,nonlocalbias,baldauf} and compute 
the non-local bias coefficients up to third order in perturbation theory. 
Our starting point is the evolution over cosmic time and in Eulerian space of the halo 
progenitors - the so-called proto-halos - until their virialization.
The basic idea is that, while  their shapes and topology change as a function of time 
(smaller substructures gradually merge to form the final halo), their centre of mass 
moves along a well-defined trajectory determined by the surrounding mass density field
\cite{max}.
Therefore, unlike virialized halos that undergo merging, by construction proto-halos
always preserve their identity. Their total number is therefore conserved over time, 
such that we can write a continuity equation for their number density

\be
\dot{\delta}_{\rm h}(\vx,\t)+\bnabla\cdot\left[(1+\delta_\h(\vx,\t))\vv(\vx,\t)\right]=0.
\ee
We may subtract from it  the  DM mass conservation equation of motion

\be
\dot{\delta}(\vx,\t)+\bnabla\cdot\left[(1+\delta(\vx,\t))\vv(\vx,\t)\right]=0,
\ee
where we have assumed unbiased halo velocity (we will relax this assumption later), 
to get

\be
\label{fund}
\dot{\delta}_{\rm h}(\vx,\t)-\dot{\delta}(\vx,\t)
+\bnabla\cdot\left[(\delta_\h(\vx,\t)-\delta(\vx,\t))\vv(\vx,\t)\right]=0.
\ee
This is the fundamental equation which we will solve order by order in perturbation theory, 
and which gives rise to a non-local bias expansion. For the sake of simplicity, we will 
restrict ourselves to a matter-dominated Universe and denote by $\bx$ the comoving spatial 
coordinates, $\tau=\int \d t/a$  the conformal time where $a$ the scale factor in the FRW 
metric, and ${\cal{H}}=\d\ln a/\d\t$  the conformal expansion rate.  
In addition, $\delta(\bx,\t)=(\rho(\bx,\t)/\overline{\rho}-1)$ is the overdensity over the 
mean matter density $\overline{\rho}$, $\delta_{\rm h}(\bx,\t)$ stands for the halo counterpart, 
and $\bv(\bx,\t)$ is the common peculiar velocity. 
In the following we will also denote by 
$\Phi(\bx,\t)$ the gravitational potential induced by density fluctuations. 
We begin with a review of the first- and second-order calculations before deriving the 
third-order nonlocal biases.

\subsection{First-order}
At first-order from Eq. (\ref{fund}) we get

\be
\dot{\delta}^\1_{\rm h}(\vx,\t)-\dot{\delta}^\1(\vx,\t)=0,
\ee
or

\be
\delta^\1_\h(\vx,\t)=\delta^\1_\h(\vx,\tau_i)+\delta^\1(\vx,\t)-\delta^\1(\vx,\tau_i),
\ee
where $\tau_i$ is some initial time. 
We assume that the initial bias expansion is local and depends only on the linear DM density 
contrast through the (Lagrangian) bias coefficients

\begin{eqnarray}
\delta_\h(\vx,\tau_i)&=&\sum_\ell\frac{b^{\rm L}_\ell(\t_i)}{\ell!}\left(\delta^\1(\vx,\tau_i)\right)^\ell\nonumber\\
&=&\sum_\ell\frac{b^{\rm L}_\ell(\t)}{\ell!}\left(\delta^\1(\vx,\tau)\right)^\ell\nonumber\\
&\simeq&
b^{\rm L}_1(\tau)\delta^\1(\vx,\tau)+\frac{1}{2}b^{\rm L}_2(\t)\left(\delta^\1(\vx,\tau)\right)^2+\nonumber\\
&&\frac{1}{3!}b^{\rm L}_3(\t)\left(\delta^\1(\vx,\tau)\right)^3+\cdots,
\end{eqnarray}
where $b^{\rm L}_\ell(\tau)=b^{\rm L}_\ell(\tau_i)(a(\t_i)/a(\t))^\ell$.  
Using $\delta^\1_\h(\vx,\tau_i)=b^{\rm L}_1(\tau)\delta(\vx,\tau)$, we obtain the standard result

\be
\label{res1}
\delta^\1_\h(\vx,\t)\simeq \left(1+b^{\rm L}_1(\tau)\right)\delta^\1(\vx,\t).
\ee
\subsection{Second-order}
At second-order we may use the first-order  result to write Eq. (\ref{fund}) in the form

\begin{eqnarray}
\dot{\delta}^\2_{\rm h}(\vx,\t)&-&\dot{\delta}^\2(\vx,\t)\nonumber\\
&+&\bnabla\cdot\left[(\delta^\1_\h(\vx,\t)-\delta^\1(\vx,\t))\vv^\1(\vx,\t)\right]=0,\nonumber\\
\end{eqnarray}
which is solved by
{\allowdisplaybreaks
\begin{eqnarray}
\delta^\2_\h(\vx,\t)&=&\delta^\2_\h(\vx,\tau_i)+
\delta^\2(\vx,\t)
-\int^\tau\d\eta\,b^{\rm L}_1(\eta)\bnabla\cdot\left[\delta^\1(\vx,\eta)\vv^\1(\vx,\eta)\right]\nonumber\\
&\simeq&\frac{1}{2}b^{\rm L}_2(\tau)\left(\delta^\1(\vx,\tau)\right)^2+\delta^\2(\vx,\t)
-\int^\tau\d\eta\,b^{\rm L}_1(\eta)\bnabla\cdot\left[\delta^\1(\vx,\eta)\vv^\1(\vx,\eta)\right]\nonumber\\
&=&\frac{1}{2}b^{\rm L}_2(\tau)\left(\delta^\1(\vx,\tau)\right)^2+\delta^\2(\vx,\t)
-
\int^\tau\d\eta\,b^{\rm L}_1(\eta)\bnabla\delta^\1(\vx,\eta)\cdot\vv^\1(\vx,\eta)
+
\int^\tau\d\eta\,b^{\rm L}_1(\eta)\delta^\1(\vx,\eta)\dot{\delta}^\1(\vx,\eta)\nonumber\\
&=&\frac{1}{2}b^{\rm L}_2(\tau)\left(\delta^\1(\vx,\tau)\right)^2+\delta^\2(\vx,\t)
-\frac{\tau}{2}b^{\rm L}_1(\tau)
\bnabla\delta^\1(\vx,\tau)\cdot\vv^\1(\vx,\tau)
+
b^{\rm L}_1(\tau)\left(\delta^\1(\vx,\tau)\right)^2,
\end{eqnarray}
}
or

\begin{eqnarray}
\delta^\2_\h(\vx,\t)&=&-\frac{1}{{\cal H}}b^{\rm L}_1(\tau)
\bnabla\delta^\1(\vx,\tau)\cdot\vv^\1(\vx,\tau)\nonumber\\
&+&\frac{1}{2}b^{\rm L}_2(\tau)\left(\delta^\1(\vx,\tau)\right)^2+\delta^\2(\vx,\t)
+
b^{\rm L}_1(\tau)\left(\delta^\1(\vx,\tau)\right)^2.\nonumber\\
&&
\end{eqnarray}
To perform the time integrals, we have used the scalings provided by the first-order quantities 
(in matter-domination)

\begin{eqnarray}
v_i^\1(\vx,\tau)&=&-\frac{\tau}{3}\partial_i\varphi(\vx)=-\frac{2}{3{\cal H}}\partial_i\varphi(\vx),\nonumber\\
\delta^\1(\vx,\tau)&=&\frac{\tau^2}{6}\nabla^2\varphi(\vx)=\frac{2}{3{\cal H}^2}\nabla^2\varphi(\vx),
\end{eqnarray}
where  $\varphi(\vx)$ is the initial condition for the gravitational potential $\Phi(\vx,\t)$. 
In order to elaborate further the second-order halo density contrast, we remind the reader that \cite{peebles,bartolo}

\begin{eqnarray}
 \delta^\2(\vx,\tau)&=&\frac{\tau^4}{2\cdot 126}\left[5(\nabla^2\varphi(\vx))^2+2\partial_k\partial_p\varphi(\vx)\partial^k\partial^p\varphi(\vx)+7\partial^i\varphi(\vx)\nabla^2\partial_i\varphi(\vx)
 \right]\nonumber\\
 &=&\frac{5}{7}\left(\delta^\1(\vx,\tau)\right)^2+\frac{8}{63{\cal H}^4}\partial_k\partial_p\varphi(\vx)\partial^k\partial^p\varphi(\vx)-\frac{1}{{\cal H}}\bnabla\delta^\1(\vx,\tau)\cdot\vv^\1(\vx,\tau)
\end{eqnarray}
and define the non-local bias operator
 
 \be
 s_{ij}(\vx,\t)=\frac{2}{3{\cal H}^2}\partial_i\partial_j\Phi(\vx,\tau)-\frac{1}{3}\delta_{ij}\delta(\vx,\tau),
 \ee
to get 

\be
\partial_k\partial_p\varphi(\vx)\partial^k\partial^p\varphi(\vx)=\frac{9{\cal H}^4}{4}\left(s^\1(\vx,\tau)\right)^2+\frac{3{\cal H}^4}{4}\left(\delta^\1(\vx,\tau)\right)^2.
\ee 
 From this expression  we deduce
 
  \begin{eqnarray}
  \label{zu}
 \delta^\2(\vx,\tau)
 &=&\frac{17}{21}\left(\delta^\1(\vx,\tau)\right)^2+\frac{2}{7}\left(s^\1(\vx,\tau)\right)^2-\frac{1}{{\cal H}}\bnabla\delta^\1(\vx,\tau)\cdot\vv^\1(\vx,\tau).
 \end{eqnarray} 
 We finally arrive at
 
 \be
 \label{res2}
\delta^\2_\h(\vx,\t)\simeq\left(1+b^{\rm L}_1(\tau)\right)\delta^\2(\vx,\tau)+\left(\frac{1}{2}b^{\rm L}_2(\t)+\frac{4}{21}b^{\rm L}_1(\tau)\right)\left(\delta^\1(\vx,\tau)\right)^2-\frac{2}{7}b^{\rm L}_1(\tau)\left(s^\1(\vx,\tau)\right)^2.
\ee 
This result reproduces exactly the one derived in Refs. \cite{catelan98,baldauf} and shows that at second--order in perturbation theory the halo overdensity is not a local function of the underlying matter overdensity.

\subsection{Third-order}
We now proceed to the original part of the computation at third-order.   The equation to solve is 
 
 \begin{eqnarray}
 \label{qw}
\dot{\delta}^\3_{\rm h}(\vx,\t)-\dot{\delta}^\3(\vx,\t)+\bnabla\cdot\left[(\delta^\1_\h(\vx,\t)-\delta^\1(\vx,\t))\vv^\2(\vx,\t)\right]+\bnabla\cdot\left[(\delta^\2_\h(\vx,\t)-\delta^\2(\vx,\t))\vv^\1(\vx,\t)\right]&=&0,\nonumber\\
&&
\end{eqnarray}
 where \cite{bartolo}
 
 {\allowdisplaybreaks
\begin{eqnarray}
 \vv_i^\2(\vx,\tau)&=&\frac{\tau^3}{18}\left[-\partial_i\partial_j\varphi(\vx)\partial^j\varphi(\vx)-\frac{3}{7}\frac{\partial_i}{\nabla^2}\left((\nabla^2\varphi(\vx))^2-\partial_k\partial_p\varphi(\vx)\partial^k\partial^p\varphi(\vx)
\right)\right]\nonumber\\
\bnabla\cdot \vv_i^\2(\vx,\tau)&=&\frac{\tau^3}{9}\left[-\partial_j\nabla^2\varphi(\vx)\partial^j\varphi(\vx)
-\partial_i\partial_j\varphi(\vx)\partial^i\partial^j\varphi(\vx)-\frac{3}{7}\left((\nabla^2\varphi(\vx))^2-\partial_k\partial_p\varphi(\vx)\partial^k\partial^p\varphi(\vx)
\right)\right]\nonumber\\
&=&\frac{\tau^3}{18}\left[-\partial_j\nabla^2\varphi(\vx)\partial^j\varphi(\vx)
-\frac{4}{7}\partial_i\partial_j\varphi(\vx)\partial^i\partial^j\varphi(\vx)-\frac{3}{7}(\nabla^2\varphi(\vx))^2\right]. \end{eqnarray}
 Eq. (\ref{qw}) gives

 \begin{eqnarray}
 \delta^\3_\h(\vx,\t)&=&\delta^\3_\h(\vx,\tau_i)+\delta^\3(\vx,\t)\nonumber\\
&-&\int^\tau\d\eta\,b^{\rm L}_1(\eta)\bnabla\cdot\left[\delta^\1(\vx,\eta)\vv^\2(\vx,\eta)\right]\nonumber\\
&-&\int^\tau\d\eta\,b^{\rm L}_1(\eta)\bnabla\cdot\left[\delta^\2(\vx,\eta)\vv^\1(\vx,\eta)\right]\nonumber\\
&-&\int^\tau\,\d\eta\,\bnabla\cdot\left[\left(\left(\frac{1}{2}b^{\rm L}_2(\eta)+\frac{4}{21}b^{\rm L}_1(\eta)\right)\left(\delta^\1(\vx,\eta)\right)^2-\frac{2}{7}b^{\rm L}_1(\tau)\left(s^\1(\vx,\tau)\right)^2\right)\vv^\1(\vx,\eta)\right],\nonumber\\
&&
\end{eqnarray}

 or
 
 \begin{eqnarray}
 \delta^\3_\h(\vx,\t)&=&\frac{1}{3!}b^{\rm L}_3(\tau)\left(\delta^\1(\vx,\tau)\right)^3+\delta^\3(\vx,\t)\nonumber\\
&-&\int^\tau\d\eta\,b^{\rm L}_1(\eta)\bnabla\delta^\1(\vx,\eta)\cdot\vv^\2(\vx,\eta)\nonumber\\
&-&\int^\tau\d\eta\,b^{\rm L}_1(\eta)\delta^\1(\vx,\eta)\bnabla\cdot\vv^\2(\vx,\eta)\nonumber\\
&-&\int^\tau\d\eta\,b^{\rm L}_1(\eta)\bnabla\delta^\2(\vx,\eta)\cdot\vv^\1(\vx,\eta)\nonumber\\
&-&\int^\tau\d\eta\,b^{\rm L}_1(\eta)\delta^\2(\vx,\eta)\bnabla\cdot\vv^\1(\vx,\eta)\nonumber\\
&-&\int^\tau\,\d\eta\,\left(\frac{1}{2}b^{\rm L}_2(\eta)+\frac{4}{21}b^{\rm L}_1(\eta)\right)\bnabla\left[\left(\delta^\1(\vx,\eta)\right)^2\right]\cdot
\vv^\1(\vx,\eta)\nonumber\\
&-&\int^\tau\,\d\eta\,\left(\frac{1}{2}b^{\rm L}_2(\eta)+\frac{4}{21}b^{\rm L}_1(\eta)\right)\left(\delta^\1(\vx,\eta)\right)^2
\bnabla\cdot\vv^\1(\vx,\eta)\nonumber\\
&+&\frac{2}{7}\int^\tau\,\d\eta\,b^{\rm L}_1(\eta)\bnabla \left(s^\1(\vx,\eta)\right)^2\cdot \vv^\1(\vx,\eta)\nonumber\\
&+&\frac{2}{7}\int^\tau\,\d\eta\,b^{\rm L}_1(\eta) \left(s^\1(\vx,\eta)\right)^2\bnabla\cdot \vv^\1(\vx,\eta).
\end{eqnarray}

 We can integrate over time to obtain

 \begin{eqnarray}
 \label{tent}
 \delta^\3_\h(\vx,\t)&=&\frac{1}{3!}b^{\rm L}_3(\tau)\left(\delta^\1(\vx,\tau)\right)^3+\delta^\3(\vx,\t)\nonumber\\
&-&\frac{\tau}{4}b^{\rm L}_1(\tau)\bnabla\delta^\1(\vx,\tau)\cdot\vv^\2(\vx,\tau)\nonumber\\
&-& \frac{\tau}{4}  b^{\rm L}_1(\tau)\delta^\1(\vx,\tau)\bnabla\cdot\vv^\2(\vx,\tau)\nonumber\\
&-& \frac{\tau}{4}b^{\rm L}_1(\tau)\bnabla\delta^\2(\vx,\tau)\cdot\vv^\1(\vx,\tau)\nonumber\\
&-& \frac{\tau}{4}b^{\rm L}_1(\tau)\delta^\2(\vx,\tau)\bnabla\cdot\vv^\1(\vx,\tau)\nonumber\\
&-& \left(\frac{\tau}{4}b^{\rm L}_2(\tau)+\frac{\tau}{21}b^{\rm L}_1(\tau)\right)\left\{\bnabla\left[\left(\delta^\1(\vx,\tau)\right)^2\right]\cdot
\vv^\1(\vx,\tau)+\left(\delta^\1(\vx,\tau)\right)^2
\bnabla\cdot\vv^\1(\vx,\tau)\right\}\nonumber\\
&+&\frac{\tau}{14}b^{\rm L}_1(\tau)\bnabla \left(s^\1(\vx,\tau)\right)^2\cdot \vv^\1(\vx,\tau)\nonumber\\
&+&\frac{\tau}{14}b^{\rm L}_1(\tau) \left(s^\1(\vx,\tau)\right)^2\bnabla\cdot \vv^\1(\vx,\tau).
\end{eqnarray}
}
Now, the mass conservation equation for the DM at third-order reads

\begin{eqnarray}
\dot{\delta}^\3(\vx,\t)+\bnabla\cdot\vv^\3(\vx,\t)&=&-\bnabla\delta^\2(\vx,\t)\cdot\vv^\1(\vx,\t)
-\bnabla\delta^\1(\vx,\t)\cdot\vv^\2(\vx,\t)\nonumber\\
&-&\delta^\2(\vx,\t)\bnabla\cdot\vv^\1(\vx,\t)-\delta^\1(\vx,\t)\bnabla\cdot\vv^\2(\vx,\t).\nonumber\\
\end{eqnarray}
Since ${\delta}^\3(\vx,\t)$ scales like $a^3$,   we can rewrite it  as

 \begin{eqnarray}
3{\cal H}{\delta}^\3(\vx,\t)+\bnabla\cdot\vv^\3(\vx,\t)&=&-\left[\bnabla\delta^\2(\vx,\t)\cdot\vv^\1(\vx,\t)
+\bnabla\delta^\1(\vx,\t)\cdot\vv^\2(\vx,\t)\right.\nonumber\\
&+&\left.\delta^\2(\vx,\t)\bnabla\cdot\vv^\1(\vx,\t)+\delta^\1(\vx,\t)\bnabla\cdot\vv^\2(\vx,\t)\right].\nonumber\\
\end{eqnarray}
Eq. (\ref{tent}) then becomes

 {\allowdisplaybreaks
 \begin{eqnarray}
 \label{oo}
 \delta^\3_\h(\vx,\t)&=&\frac{1}{3!}b^{\rm L}_3(\tau)\left(\delta^\1(\vx,\tau)\right)^3+\left(\frac{3}{2}b^{\rm L}_1(\tau)+1\right)\delta^\3(\vx,\t)+
 \frac{1}{2\cal H}b^{\rm L}_1(\tau)\theta^\3(\vx,\t)
 \nonumber\\
&-& \left(\frac{\tau}{4}b^{\rm L}_2(\tau)+\frac{\tau}{21}b^{\rm L}_1(\tau)\right)\left\{\bnabla\left[\left(\delta^\1(\vx,\tau)\right)^2\right]\cdot
\vv^\1(\vx,\tau)+\left(\delta^\1(\vx,\tau)\right)^2
\bnabla\cdot\vv^\1(\vx,\tau)\right\}\nonumber\\
&+&\frac{\tau}{14}b^{\rm L}_1(\tau)\bnabla \left(s^\1(\vx,\tau)\right)^2\cdot \vv^\1(\vx,\tau)\nonumber\\
&+&\frac{\tau}{14}b^{\rm L}_1(\tau) \left(s^\1(\vx,\tau)\right)^2\bnabla\cdot \vv^\1(\vx,\tau),
\end{eqnarray}
}
where  $\theta(\vx,\t)=\bnabla\cdot\vv(\vx,\t)$ satisfies  at any order in perturbation theory the DM momentum equation 
 
\begin{eqnarray}
\dot{\theta}(\vx,\t)+{\cal H}\theta(\vx,\t)+\partial^j v_i(\vx,\t)\partial^i v_j(\vx,\t)+\vv(\vx,\t)\cdot\bnabla\theta(\vx,\t)=-\frac{3}{2}{\cal H}^2\delta(\vx,\t).
\end{eqnarray}
Following Refs. \cite{MD,norena}, we introduce another non-local coefficient
 
 \be
 t_{ij}(\vx,\t)=
 -\frac{1}{{\cal H}}\left(\partial_iv_j(\vx,\t)-\frac{1}{3}\delta_{ij}\theta(\vx,\t)\right)-s_{ij}(\vx,\t).
 \ee
 It is traceless and vanishes at first-order in perturbation theory
 as
 
 \be
 s^\1_{ij}(\vx,\t)=
 -\frac{1}{{\cal H}}\partial_iv^\1_j(\vx,\t)-\frac{1}{3}\delta_{ij}\delta^\1(\vx,\t),
 \ee
 and therefore 
 
  \be
 t^\1_{ij}(\vx,\t)=\frac{1}{3}\delta_{ij}\left(\frac{1}{{\cal H}}\theta^\1(\vx,\t)+\delta^\1(\vx,\t)\right)=0.
  \ee
 This implies that $t^2(\vx,\t)=t_{ij}(\vx,\t)t^{ij}(\vx,\t)$ is fourth-order and we can neglect it here and henceforth. In particular, up to third-order,
 
 \be
 \partial^j v_i(\vx,\t)\partial^i v_j(\vx,\t)=-2{\cal H}^2t(\vx,\t)\cdot s(\vx,\t)+\frac{1}{3}\theta^2(\vx,\t)+{\cal H}^2 s^2(\vx,\t).
 \ee
 The equation for $\theta(\vx,\t)$ then becomes
 
\begin{eqnarray}
\label{theta}
\dot{\theta}(\vx,\t)+{\cal H}\theta(\vx,\t)-2{\cal H}^2 t(\vx,\t)\cdot s(\vx,\t)+\frac{1}{3}\theta^2(\vx,\t)+{\cal H}^2 s^2(\vx,\t)
+\vv(\vx,\t)\cdot\bnabla\theta(\vx,\t)=-\frac{3}{2}{\cal H}^2\delta(\vx,\t).\nonumber\\
&&
\end{eqnarray}
%
Since

\begin{eqnarray}
\bnabla\cdot \vv^\2(\vx,\tau)
&=&\frac{\tau^3}{18}\left[-\partial_j\nabla^2\varphi(\vx)\partial^j\varphi(\vx)
-\frac{4}{7}\partial_i\partial_j\varphi(\vx)\partial^i\partial^j\varphi(\vx)-\frac{3}{7}(\nabla^2\varphi(\vx))^2\right]\nonumber\\
&=&\bnabla\delta^\1(\vx,\tau)\cdot \vv^\1(\vx,\t)-\frac{4}{7}{\cal H}\left(s^\1(\vx,\tau)\right)^2-\frac{13}{21}{\cal H}\left(\delta^\1(\vx,\t)\right)^2,
\end{eqnarray} 
we have

\be
-\frac{1}{\cal H}\bnabla\cdot \vv^\2(\vx,\tau)-\delta^\2(\vx,\t)=\frac{2}{7}\left(s^\1(\vx,\tau)\right)^2-\frac{4}{21}\left(\delta^\1(\vx,\t)\right)^2,
\ee
 and it is convenient to define  two new non-local bias operators \cite{MD}
 
 \be
 \eta(\vx,\t)=-\frac{\theta(\vx,\tau)}{\cal H}-\delta(\vx,\t)
 \ee
 and
 \be
 \psi(\vx,\t)=\eta(\vx,\t)-\frac{2}{7}s^2(\vx,\tau)+\frac{4}{21}\delta^2(\vx,\t).
\ee
 By construction $\eta(\vx,\t)$ is a second-order quantity and $\psi(\vx,\t)$ is third-order quantity.
 Eq. (\ref{oo}) becomes then
  {\allowdisplaybreaks
 \begin{eqnarray}
 \delta^\3_\h(\vx,\t)&=&\frac{1}{3!}b^{\rm L}_3(\tau)\left(\delta^\1(\vx,\tau)\right)^3+\left(\frac{3}{2}b^{\rm L}_1(\tau)+1\right)\delta^\3(\vx,\t)\nonumber\\
 &+&\frac{1}{2} b^{\rm L}_1(\tau)\left(-\psi^\3(\vx,\t) -\delta^\3(\vx,\t)-\frac{4}{7} s^\1(\vx,\tau)s^\2(\vx,\tau) +\frac{8}{21}\delta^\1(\vx,\t)\delta^\2(\vx,\t) \right)
 \nonumber\\
&-&\frac{1}{{\cal H}} \left(\frac{1}{2}b^{\rm L}_2(\tau)+\frac{2}{21}b^{\rm L}_1(\tau)\right)\left\{\bnabla\left[\left(\delta^\1(\vx,\tau)\right)^2\right]\cdot
\vv^\1(\vx,\tau)+\left(\delta^\1(\vx,\tau)\right)^2
\bnabla\cdot\vv^\1(\vx,\tau)\right\}\nonumber\\
&+&\frac{1}{7{\cal H}}b^{\rm L}_1(\tau)\bnabla \left(s^\1(\vx,\tau)\right)^2\cdot \vv^\1(\vx,\tau)\nonumber\\
&+&\frac{1}{7{\cal H}}b^{\rm L}_1(\tau) \left(s^\1(\vx,\tau)\right)^2\bnabla\cdot \vv^\1(\vx,\tau).
\end{eqnarray}
Using Eq. (\ref{zu}) we can  rewrite it as 

\begin{eqnarray}
 \delta^\3_\h(\vx,\t)&=&\frac{1}{3!}b^{\rm L}_3(\tau)\left(\delta^\1(\vx,\tau)\right)^3+(1+b^{\rm L}_1(\tau))\delta^\3(\vx,\t)\nonumber\\
 &+&\frac{1}{2} b^{\rm L}_1(\tau)\left(-\psi^\3(\vx,\t) -\frac{4}{7} s^\1(\vx,\tau)\cdot s^\2(\vx,\tau) +\frac{8}{21}\delta^\1(\vx,\t)\delta^\2(\vx,\t) \right)
 \nonumber\\
&-&\frac{1}{{\cal H}} \left(\frac{1}{2}b^{\rm L}_2(\tau)+\frac{2}{21}b^{\rm L}_1(\tau)\right)\delta^\1(\vx,\tau)\left\{2\bnabla\delta^\1(\vx,\tau)\cdot
\vv^\1(\vx,\tau)-{\cal H}\left(\delta^\1(\vx,\tau)\right)^2\right\}\nonumber\\
&+&\frac{1}{7{\cal H}}b^{\rm L}_1(\tau)\bnabla \left(s^\1(\vx,\tau)\right)^2\cdot \vv^\1(\vx,\tau)\nonumber\\
&+&\frac{1}{7{\cal H}}b^{\rm L}_1(\tau) \left(s^\1(\vx,\tau)\right)^2\bnabla\cdot \vv^\1(\vx,\tau),
\end{eqnarray}
or
\begin{eqnarray}
\label{df}
 \delta^\3_\h(\vx,\t)
&=&(1+b^{\rm L}_1(\tau))\delta^\3(\vx,\t)
\nonumber\\
&+&2 \left(\frac{1}{2}b^{\rm L}_2(\tau)+\frac{4}{21}b^{\rm L}_1(\tau)\right)\delta^\1(\vx,\tau)\delta^\2(\vx,\tau)
\nonumber\\
&+&\left[\frac{1}{3!}b^{\rm L}_3(\tau)-\frac{13}{21}\left(\frac{1}{2}b^{\rm L}_2(\tau)+\frac{2}{21}b^{\rm L}_1(\tau)\right)\right]\left(\delta^\1(\vx,\tau)\right)^3\nonumber\\
 &+&\frac{1}{2} b^{\rm L}_1(\tau)\left(-\psi^\3(\vx,\t) -\frac{4}{7} s^\1(\vx,\tau)\cdot s^\2(\vx,\tau)\right)
\nonumber\\
&-& \frac{4}{7}\left(\frac{1}{2}b^{\rm L}_2(\tau)+\frac{29}{84}b^{\rm L}_1(\tau)\right)\delta^\1(\vx,\tau)\left(s^\1(\vx,\tau)\right)^2\nonumber\\
&+&\frac{1}{7{\cal H}}b^{\rm L}_1(\tau)\bnabla \left(s^\1(\vx,\tau)\right)^2\cdot \vv^\1(\vx,\tau).
\end{eqnarray}
Let us concentrate on the last term of the expression (\ref{df}). We can operate a series of manipulations on it

\begin{eqnarray}
\bnabla \left(s^\1(\vx,\tau)\right)^2\cdot \vv^\1(\vx,\tau)&=&2 s_{ij}^\1(\vx,\t)v_k^\1(\vx,\t)\del^k s_{ij}^\1(\vx,\t)\nonumber\\
&=&-\frac{2}{{\cal H}} s_{ij}^\1(\vx,\t)v_k^\1(\vx,\t)\del^k\del_i v_j^\1(\vx,\t) \nonumber \\
&=&-\frac{2}{{\cal H}} s_{ij}^\1(\vx,\t)\left[\del_i\left(v_k^\1(\vx,\t)\del^k v_j^\1\ar\right) -\del_i v_k^\1(\vx,\t)\del^k v_j^\1\ar \right]\nonumber \\
&=&-\frac{2}{{\cal H}}s_{ij}^\1(\vx,\t)\left[-\del_i\del_j \Phi^\2\ar -\del_i
\dot{v}_j^\2\ar-{\cal H} \del_iv_j^\2\ar\right]\nonumber \\
&&+2{\cal H}s_{ij}^\1(\vx,\t)\left(s_{ik}^\1\ar+\frac{1}{3}\delta_{ik}\delta^\1\ar\right)\left(s_{kj}^\1\ar+\frac{1}{3}\delta_{kj}\delta^\1\ar\right)\nonumber \\
&=&-\frac{2}{{\cal H}} s_{ij}^\1(\vx,\t)\left(-\frac{3{\cal H}^2}{2}s_{ij}^\2\ar\right)-\frac{2}{{\cal H}} s_{ij}^\1(\vx,\t)\left(-\frac{5}{2}{\cal H}\del_iv_j^\2\ar\right)\nonumber \\
&&+2{\cal H}s_{ij}^\1(\vx,\t)\left(s_{ik}^\1\ar s_{kj}^\1\ar +\frac{2}{3}s_{ij}^\1 \ar \delta^\1\ar\right)\nonumber \\
&=&3{\cal H}s_{ij}^\1(\vx,\t)s_{ij}^\2(\vx,\t)+5 s_{ij}^\1(\vx,\t){\cal H}\left(
\frac{1}{3{\cal H}}\delta_{ij}\theta^\2\ar-t_{ij}^\2\ar-s_{ij}^\2\ar\right)\nonumber \\
&&+2 {\cal H} \left(s^\1\ar\right)^3+\frac{4}{3}{\cal H}\left(s^\1\ar\right)^2
\delta^\1\ar\nonumber \\
&=&-2{\cal H}s_{ij}^\1(\vx,\t)s_{ij}^\2(\vx,\t)-5 {\cal H}s_{ij}^\1(\vx,\t)t_{ij}^\2(\vx,\t)\nonumber\\
&+&2 {\cal H} \left(s^\1\ar\right)^3+\frac{4}{3}{\cal H}\left(s^\1\ar\right)^2
\delta^\1\ar.
\end{eqnarray}
Therefore
\begin{eqnarray}
\frac{1}{7{\cal H}}b^{\rm L}_1(\t) \bnabla \left(s^\1(\vx,\tau)\right)^2\cdot \vv^\1(\vx,\tau)&=&-\frac{2}{7}b^{\rm L}_1(\t)s_{ij}^\1(\vx,\t)s_{ij}^\2(\vx,\t)-\frac{5}{7}b^{\rm L}_1(\t) s_{ij}^\1(\vx,\t)t_{ij}^\2(\vx,\t)\nonumber \\
&&+\frac{2}{7}b^{\rm L}_1(\t)  \left(s^\1\ar\right)^3+\frac{4}{21}b^{\rm L}_1(\t)\left(s^\1\ar\right)^2
\delta^\1\ar, \nonumber\\
&&
\end{eqnarray}
so that 

\begin{eqnarray}
\label{res3}
 \delta^\3_\h(\vx,\t)
&=&(1+b^{\rm L}_1(\tau))\delta^\3(\vx,\t)
\nonumber\\
&+&2 \left(\frac{1}{2}b^{\rm L}_2(\tau)+\frac{4}{21}b^{\rm L}_1(\tau)\right)\delta^\1(\vx,\tau)\delta^\2(\vx,\tau)
\nonumber\\
&+&\left[\frac{1}{3!}b^{\rm L}_3(\tau)-\frac{13}{21}\left(\frac{1}{2}b^{\rm L}_2(\tau)+\frac{2}{21}b^{\rm L}_1(\tau)\right)\right]\left(\delta^\1(\vx,\tau)\right)^3\nonumber\\
 &+&\frac{1}{2} b^{\rm L}_1(\tau)\left(-\psi^\3(\vx,\t) -\frac{8}{7} s^\1(\vx,\tau)\cdot s^\2(\vx,\tau)\right)
\nonumber\\
&-& \frac{4}{7}\left(\frac{1}{2}b^{\rm L}_2(\tau)+\frac{2}{81}b^{\rm L}_1(\tau)\right)\delta^\1(\vx,\tau)\left(s^\1(\vx,\tau)\right)^2\nonumber\\
&-&\frac{5}{7}b^{\rm L}_1(\t) s_{ij}^\1(\vx,\t)t_{ij}^\2(\vx,\t)+\frac{2}{7}b^{\rm L}_1(\t)  \left(s^\1\ar\right)^3.
\end{eqnarray}
Combining the results (\ref{res1}),  (\ref{res2}), and  (\ref{res3}), we finally get

\begin{eqnarray}
\delta_\h(\vx,\t)&=&(1+b^{\rm L}_1(\tau))\delta(\vx,\t)+\left(\frac{1}{2}b^{\rm L}_2(\t)+\frac{4}{21}b^{\rm L}_1(\tau)\right)\delta^2(\vx,\tau)-\frac{2}{7}b^{\rm L}_1(\tau)s^2(\vx,\tau)\nonumber\\
&+&\left[\frac{1}{3!}b^{\rm L}_3(\tau)-\frac{13}{21}\left(\frac{1}{2}b^{\rm L}_2(\tau)+\frac{2}{21}b^{\rm L}_1(\tau)\right)\right]\delta^3(\vx,\tau)\nonumber\\
&-&\frac{1}{2}b^{\rm L}_1(\tau)\psi- \frac{4}{7}\left(\frac{1}{2}b^{\rm L}_2(\tau)+\frac{2}{81}b^{\rm L}_1(\tau)\right)\delta(\vx,\tau)s^2(\vx,\tau)\nonumber\\
&-&\frac{5}{7}b^{\rm L}_1(\t) s(\vx,\t)\cdot t(\vx,\t)+\frac{2}{7}b^{\rm L}_1(\t) s^3(\vx,\t).
\end{eqnarray}
}
The halo density contrast can thus be written as

\begin{eqnarray}\label{eq:halo}
\delta_{\rm h}\ar&=&b_1\delta\ar+\frac{1}{2!}b_2\delta^2\ar+
\frac{1}{3!}b_3\delta^3\ar+\frac{1}{2}b_{s^2} s^2\ar+b_\psi\psi\ar+b_{st}s\ar\cdot t\ar+\cdots,\nonumber\\
&&
\end{eqnarray}
where the relevant bias coefficients are 

\be\label{eq:biasparameters}
 \fbox{$\displaystyle
b_1=1+b^{\rm L}_1,\,\,
b_2=b^{\rm L}_2+\frac{8}{21}b^{\rm L}_1,\,\,
b_{s^2}=-\frac{4}{7}b^{\rm L}_1,\,\,
b_\psi=-\frac{1}{2}b^{\rm L}_1,\,\,
b_{st}=-\frac{5}{7}b^{\rm L}_1$}.
\ee
This is the main result of this Section. 

\subsection{Scale-dependent bias}
\noindent
So far, we have not considered the possibility that the Lagrangian bias factors may be scale-dependent. 
In peak theory for instance, the peak constraint induces $k$-dependent corrections at all orders in the
bias coefficients \cite{des08,des10,des13,Verde:2014nwa}. Moreover, peak velocities are statistically biased at linear
order, and this bias propagates to higher order owing to gravity mode-coupling \cite{desjacquessheth10,des10}. 
Even though calculations within the peak formalism are relatively tedious, the modifications brought by
the peak constraint can actually be easily implemented in any Lagrangian approach \cite{des10,des13}: 
the scale-independent Lagrangian bias factors $b_n^\text{L}$ should be replaced (in Fourier space) by the 
scale-dependent functions $c_n^\text{L}(\vk_1,\dots\vk_n,\t_i)$. At the lowest orders,

\begin{align}
c_1^\text{L}(k,\t_i) &= b_{10}^\text{L}(\t_i) + b_{01}^\text{L}(\t_i) k^2, \\
c_2^\text{L}(\vk_1,\vk_2,\t_i) &=\biggl\{b_{20}^\text{L}(\t_i) + b_{11}^\text{L}(\t_i) 
\left(k_1^2+k_2^2\right) + b_{02}^\text{L}(\t_i) k_1^2 k_2^2  \nonumber \\ 
& \qquad -2 \chi_{10}^\text{L}(\t_i) \left(\vk_1\cdot\vk_2\right) +\chi_{01}^\text{L}(\t_i)
\biggl[3\left(\vk_1\cdot\vk_2\right)^2-k_1^2 k_2^2\biggr]\biggr\}.\label{eq:c2}
\end{align}
As a result, the local Lagrangian bias expansion generalizes to

\begin{align}\label{eq:newLocbias}
\delta_\h(\vx,\t_i) &= b_{10}^\text{L}(\t_i) \delta(\vx,\t_i) 
- b_{01}^\text{L}(\t_i) \bnabla^2\delta(\vx,\t_i) 
+ \frac{1}{2} b_{20}^\text{L}(\t_i)\delta^2(\vx,\t_i) \nonumber \\ 
& -  b_{11}^\text{L}(\t_i)\delta(\vx,\t_i)\bnabla^2\delta(\vx,\t_i) 
+ \frac{1}{2} b_{02}^\text{L}(\t_i)\bigl[\bnabla^2\delta(\vx,\t_i)\bigr]^2 \nonumber \\
& + \chi_{10}^\text{L}(\t_i)\bigl(\bnabla\delta\bigr)^2\!\!(\vx,\t_i) 
+ \frac{1}{2}\chi_{01}^\text{L}(\t_i)
\left[3\partial_i\partial_j\delta-\delta_{ij}\bnabla^2\delta\right]^2\!\!\!\!(\vx,\t_i) +\dots,
\end{align}
where the various bias factors $b_{ij}^\text{L}$ and $\chi_{ij}^\text{L}$ can be obtained from a 
peak-background split applied to the halo mass function (see \cite{des13} for details).

In addition, the Zel'dovich gravity mode-coupling kernels $F_n^\text{ZA}(\vk_1,\dots,\vk_n)$ should 
be replaced by 

\be\label{eq:newZAkernels}
{\cal F}_n^\text{ZA}(\vk_1,\dots,\vk_n) = 
F_n^\text{ZA}(\vk_1,\dots,\vk_n) \times b_v(k_1)\dots b_v(k_n).
\ee
Here, $b_v(k)$ is the linear velocity bias. In the peak model for instance, it is $b_v(k)=1-R_v^2 k^2$ 
where

\be\label{eq:rvpeak}
R_v^2=\frac{\sigma_0^2}{\sigma^2_1}
\ee
is a characteristic scale that is proportional to the Lagrangian radius of a halo, and 
$\sigma_j^2\equiv\left\langle k^{2j}\right\rangle$ are moments of the matter power spectrum (smoothed on the
Lagrangian halo scale). The $k^2$-dependence arises from requiring invariance under rotations. 
In fact, the linear peak velocities can be thought of as arising from the continuous, local relation 
\cite{des08,des10}

\be
\vv_\h(\vx,\t_i) = \vv(\vx,\t_i) - R_v^2 \bnabla\delta(\vx,\t_i).
\ee
This development could be generalized to include all the higher order terms consistent with the symmetry of 
the problem. However, since we are mainly interested in the first-order scale-dependent corrections that 
dominate at relatively small $k$, we will postpone such a study to future work.
Eq. (\ref{eq:newZAkernels}) reflects the fact that, in the peak approach, the linear velocity bias remains 
constant throughout time \cite{des10}. By contrast, the fluid approximation to halos and dark matter 
predicts that any initial velocity bias should decay rapidly to unity \cite{nonlocalbias}. Given that this 
discrepancy has not been resolved yet (see, however, Ref. \cite{bd14}), we will assume that the velocity bias
remains constant in order to write down expressions more general than those obtained within the fluid 
approximation.

At the first order, Eq.(\ref{res1}) generalizes easily to

\be\label{eq:peakb}
\label{eq:deltah_with_k}
\delta_\h^\1(\vx,\t) \simeq \left(1+b_{10}^\text{L}(\t)\right)\delta^\1(\vx,\t)
+\left(R_v^2 - b_{01}^\text{L}(\t)\right) \bnabla^2\delta^\1(\vx,\t),
\ee
which follows from the linearized continuity equation $\delta^\1 \propto - \bnabla\cdot\vv^\1$. This result
reproduces the findings of \cite{des10}, who explicitly took into account the peak constraint. Note also 
that $b_{01}^\text{L}(\t)=b_{01}^\text{L}(\t_i)(a(\t_i)/a(\t))$. As can be seen, the amplitude of the 
contribution proportional to $k^2$ scales as $R_v^2 - b_{01}^\text{L}(\t)$, which is generally non-zero. 
Therefore, we shall expect this $k^2$-dependence to appear at sufficiently small scales in the halo bias.

At the second order, the halo overabundance $\delta_\h(\vx,\t)$ with scale-dependent spatial and velocity 
bias can be computed by combining the results of Ref. \cite{des10} (derived in the Zel'dovich approximation) 
with those of \cite{mat} (derived at higher order in Lagrangian PT). 
The second-order halo overdensity takes the form

\begin{align}
\delta_\h(\vx,\t) &=\frac{1}{2}\int\frac{d^3q_1}{(2\pi)^3}\int\frac{d^3q_2}{(2\pi)^3}
\biggl\{{\cal F}_2^\text{ZA}(\vq_1,\vq_2)+\frac{a(\t_i)}{a(\t)}\left[
{\cal F}_1^\text{ZA}(\vq_1)c_1(q_2,\t_i)+{\cal F}_1^\text{ZA}(\vq_2)
c_1(q_1,\t_i)\right] \nonumber \\ 
& \qquad + \frac{a^2(\t_i)}{a^2(\t)}c_2(\vq_1,\vq_2,\t_i)
+\frac{2}{7}-\frac{3}{7}\left(\mu^2-\frac{1}{3}\right)\biggr\} 
\delta^\1(\vq_1,\t)\delta^\1(\vq_2,\t) \delta_\text{D}(\vk-\vq_1-\vq_2),
\label{eq:2ndOrdMC}
\end{align} 
where we have momentarily set $\delta^\1\equiv \nabla^2\phi^\1$ so that it resembles the well-known PT 
expression. The various contributions can be reduced to a combination of local and nonlocal quantities 
analogously to the calculation performed above. For instance, 
$(1/2)(a(\t_i)/a(\t))[{\cal F}_1^\text{ZA}(\vq_1)c_1(q_2,\t_i)+1\leftrightarrow 2]$ includes a term 
proportional to $(1/2)b_{01}^\text{L}(\t)\left[q_2^2 \frac{\vk\cdot\qvh_1}{q_1} +1\leftrightarrow 2\right]$ 
which, after  some manipulations, becomes

\begin{align}\label{eq:b01}
b_{01}^\text{L}(\t)\bnabla\left(\nabla^2\delta^\1\bnabla^{-1}\delta^\1\right) &=
-b_{01}^\text{L}(\t)\nabla^2\delta^\2(\vk,\t) + \frac{9}{7}b_{01}^\text{L}(\t) \delta^\1\nabla^2\delta^\1
+\frac{18}{7}b_{01}^\text{L}(\t) \left(\bnabla\delta^\1\right)^2 \nonumber \\
&\qquad + \frac{6}{7}b_{01}^\text{L}(\t)  
\left(\frac{2}{3\H^2}\partial_i\partial_j\Phi-\frac{1}{3}\delta_{ij}\delta\right)
\left(\partial_i\partial_j\delta-\frac{1}{3}\delta_{ij}\nabla^2\delta\right) \nonumber \\
&\qquad +\frac{4}{7}b_{01}^\text{L}(\t) \left(\frac{2}{3\H^2}\partial_i\partial_j\partial_k\Phi
-\frac{1}{3}\delta_{ij}\partial_k\delta\right)^2.
\end{align}
We have restored the factors of $3/(2\H^2)$ to be consistent with the units employed throughout the 
paper. 
The first term in the right-hand side adds to $-b_{01}^\text{L}\nabla^2\delta^{(1)}$ in 
Eq.(\ref{eq:deltah_with_k}) to yield $-b_{01}^\text{L}\nabla^2\delta$, where $\delta$ is the mass density
fluctuations at second order.
The following two terms contribute to the Eulerian biases $b_{11}$ and $b_{02}$. The last two terms are 
new nonlocal bias contributions, which are however heavily suppressed relative to $s^2(\vx,\t)$ as they 
carry additional powers of $\vk$. 
The remaining terms in the right-hand side of Eq. (\ref{eq:2ndOrdMC}) can be written analogously.
For the purpose of the present work however, Eq. (\ref{eq:deltah_with_k}) and its $k$-dependent contribution
at linear order are sufficient. Therefore, we generalize Eq.(\ref{eq:halo}) to

\begin{eqnarray}\label{eq:halok}
\delta_{\rm h}\ar&=&b_{10}\delta\ar - b_{01}\nabla^2\delta\ar +\frac{1}{2!}b_{20}\delta^2\ar 
\nonumber \\
&& \qquad +\frac{1}{2}b_{s^2} s^2\ar +b_\psi\psi\ar+b_{st}s\ar\cdot t\ar+\cdots, 
\end{eqnarray}
where 
\begin{equation}\label{eq:biask}
b_{10} = 1+ b_{10}^\text{L}, \quad 
b_{01} = -R_v^2 + b_{01}^\text{L}, \quad 
b_{20}=b^{\rm L}_{20}+\frac{8}{21}b^{\rm L}_{10},\quad
b_{s^2}=-\frac{4}{7}b^{\rm L}_{10},\quad
b_\psi=-\frac{1}{2}b^{\rm L}_{10},\quad
b_{st}=-\frac{5}{7}b^{\rm L}_{10}.
\end{equation}
This is the model we will consider hereafter. As we will see shortly, a scale-dependent bias at linear order 
appears necessary to explain recent numerical measurements of halo bias with massive neutrinos.


\section{Halo bias in the presence of massive neutrinos}\label{sec:neutrinobias}
\noindent
Refs. \cite{Villaescusa-Navarro:2013pva,Castorina:2013wga,Costanzi:2013bha} investigated the impact of 
massive neutrinos on the spatial distribution of dark matter halos and galaxies using large box-size 
N-body simulations that incorporate massive neutrinos as an extra set of particles.
They found that massive neutrinos generate an additional scale-dependence in the halo power spectrum on 
midly nonlinear scales, in agreement with previous theoretical predictions \cite{saito09}. 
In this Section, we will compare the scale-dependence induced by massive neutrinos with that generated 
by gravitational mode-coupling and Lagrangian halo bias. We will show that the latter is substantially 
steeper, so that it should be relatively easy to isolate the contribution of massive neutrinos from a 
measurement of the halo power spectrum.

\subsection{Perturbative approach}
\noindent
To model the impact of massive neutrinos on the clustering of dark matter halos, 
we follow \cite{saito09,ichiki11,wong, shoji10} and assume that the latter trace the cold Dark Matter 
(CDM) plus baryons fluctuation field, with a linear growth rate suppressed in a scale-dependent way by 
the massive neutrinos.
This approximation is motivated by the smallness of the neutrino masses we consider. 
For $\sum m_\nu < 0.6$ eV, the neutrino free-streaming scale is sufficiently large that the neutrino 
perturbations remain in the linear regime up to late time. It has been shown to work well in 
Refs. \cite{Villaescusa-Navarro:2013pva,Castorina:2013wga,Costanzi:2013bha}. 
We thus write the total dark matter perturbation as a weighted sum of cold Dark Matter (we will ignore 
the effect of baryons in what follows, except for the fact that our CDM transfer function is a weighted 
sum of the baryons + CDM transfer functions) and neutrino fluctuations,

\be
\delta_{\rm m}=(1-f_\nu)\delta_{\rm c}+f_\nu\delta_\nu,
\ee
where the neutrino overdensity $\delta_\nu$ is in the linear regime, and the neutrino fraction $f_\nu$ is

\be
f_\nu = \frac{\Omega_\nu}{\Omega_c+\Omega_\nu}.
\ee
Replacing $\delta\ar$ by $\delta_c\ar$ in the right-hand side of Eq.(\ref{eq:halok}), the halo-mass
cross-power spectrum reads

\be
\label{biasmatter}
P_{\rm hm}(k)= \left(b_{10}+b_{01} k^2\right)P_{\rm cm}^{\rm NL}(k)+\Delta P_{\rm hm}(k)+P_{\rm cc}(k)I_3(k),
\ee
where 

\begin{equation}\label{eq:cm}
P_{\rm cm}^{\rm NL} 
= \frac{ P_{\rm mm}^{\rm NL}-f_{\nu} P_{\rm \nu m}}{1-f_{\nu}}
= \frac{ P_{\rm mm}^{\rm NL}-f_\nu\left(1-f_\nu\right) P_{{\rm c\nu}}-f_\nu^2 P_{{\rm \nu\nu}}}{1-f_\nu},
\end{equation}
and

{\allowdisplaybreaks
\begin{eqnarray}\label{eq:deltaphm}
\Delta P_{\rm hm}(k)
&=&(1-f_\nu)\,b_{20}
\int\frac{\d^3 q}{(2\pi)^3}P_{\rm cc}(q)P_{\rm cc}(| \vec{k}-\vec{q}|)F_2(\vq,\vk-\vq)
\nonumber\\
&+&(1-f_\nu)\,b_{s^2}
\int\frac{\d^3 q}{(2\pi)^3}P_{\rm cc}(q)P_{\rm cc}(| \vec{k}-\vec{q}|)F_2(\vq,\vk-\vq)S(\vq,\vk-\vq)
\end{eqnarray}
Here, $P_{{\rm cc}}$, $P_{{\rm \nu\nu}}$ and $P_{{\rm c\nu}}$ are the linear CDM, neutrinos power spectrum
and the CDM-neutrinos cross-power spectrum, respectively. 
We have adopted the notation of Refs. \cite{MD,mc} for our definition of
\begin{eqnarray}\label{i3}
I_3(k)&=&\frac{32}{105}(1-f_\nu)\left(b_{st}-\frac{5}{2}b_{s^2}
+\frac{16}{21}b_\psi\right)\int\d\ln r\,\Delta^2_{\rm cc}(kr)\,I_R(r),
\end{eqnarray}
being
\begin{eqnarray}
I_R(r)&=&I(r)+\frac{5}{6},\nonumber\\
I(r)&=&\frac{105}{32}\int_{-1}^1\d\mu\,D_2(\-\vq,\vk)S(\vq,\vk-\vq)
\end{eqnarray}
and
\begin{eqnarray}
F_2(\vq_1,\vq_2)&=&\frac{5}{7}+\frac{1}{2}\frac{\vq_1\cdot\vq_2}{q_1 q_2}\left(\frac{q_1}{q_2}
+\frac{q_2}{q_1}\right)+\frac{2}{7}\left(\frac{\vq_1\cdot\vq_2}{q_1 q_2}\right)^2,\nonumber\\
S(\vq_1,\vq_2)&=&\left(\frac{\vq_1\cdot\vq_2}{q_1 q_2}\right)^2-\frac{1}{3},\nonumber\\
D_2(\vq_1,\vq_2)&=&\frac{2}{7}\left[S(\vq_1,\vq_2)-\frac{2}{3}\right] .
\end{eqnarray}
The latter are the second-order perturbative kernels. Note that, below the neutrino free-streaming scale, 
the cross-power spectrum $P_\text{cm}^\text{NL}$ is enhanced by a factor of $(1-f_\nu)^{-1}$ relative to 
$P_\text{mm}^\text{NL}$. 
Following Ref. \cite{MD}, one should interpret $b_{st}$ etc. as ``renormalized'' bias parameters. 
Eq. (\ref{eq:biasparameters}) leads to the following relation

\be\label{bnl}
\fbox{$\displaystyle
\frac{32}{105}\left(b_{st}-\frac{5}{2}b_{s^2}+\frac{16}{21}b_\psi\right)=\frac{32}{315} b_{10}^{\rm L}\equiv b_{\rm NL}$}.
\ee
Further details regarding the evaluation of these integrals can be found in Appendix \S\ref{ap:hm}. 

The nonlinear mass power spectrum $P_{\rm mm}^\text{NL}$ and the linear $P_{\rm cc}$, $P_{\rm c\nu}$ and 
$P_{\nu\nu}$, Eqs. (\ref{biasmatter}),(\ref{eq:cm}) and (\ref{eq:deltaphm}), in the presence of massive 
neutrinos are obtained from the {\small CLASS} code \cite{Blas:2011rf} (see also Ref. \cite{Lesgourgues:2011re} 
for an overview). 
We refer the reader to \cite{Bird:2011rb} for details about this implementation. 

\subsection{Bias Parameters}\label{sec:biasp}
\noindent
To evaluate the halo-matter power spectrum Eq. (\ref{biasmatter}), we need predictions for the values of 
the bias parameters $b^{\rm L}_{10}$, $b^{\rm L}_{01}$, $b^{\rm L}_{20}$ together with the scale $R_v$
that quantifies the magnitude of the $k$-dependent velocity bias.

\begin{figure*}
\centering 
\resizebox{0.6\textwidth}{!}{\includegraphics{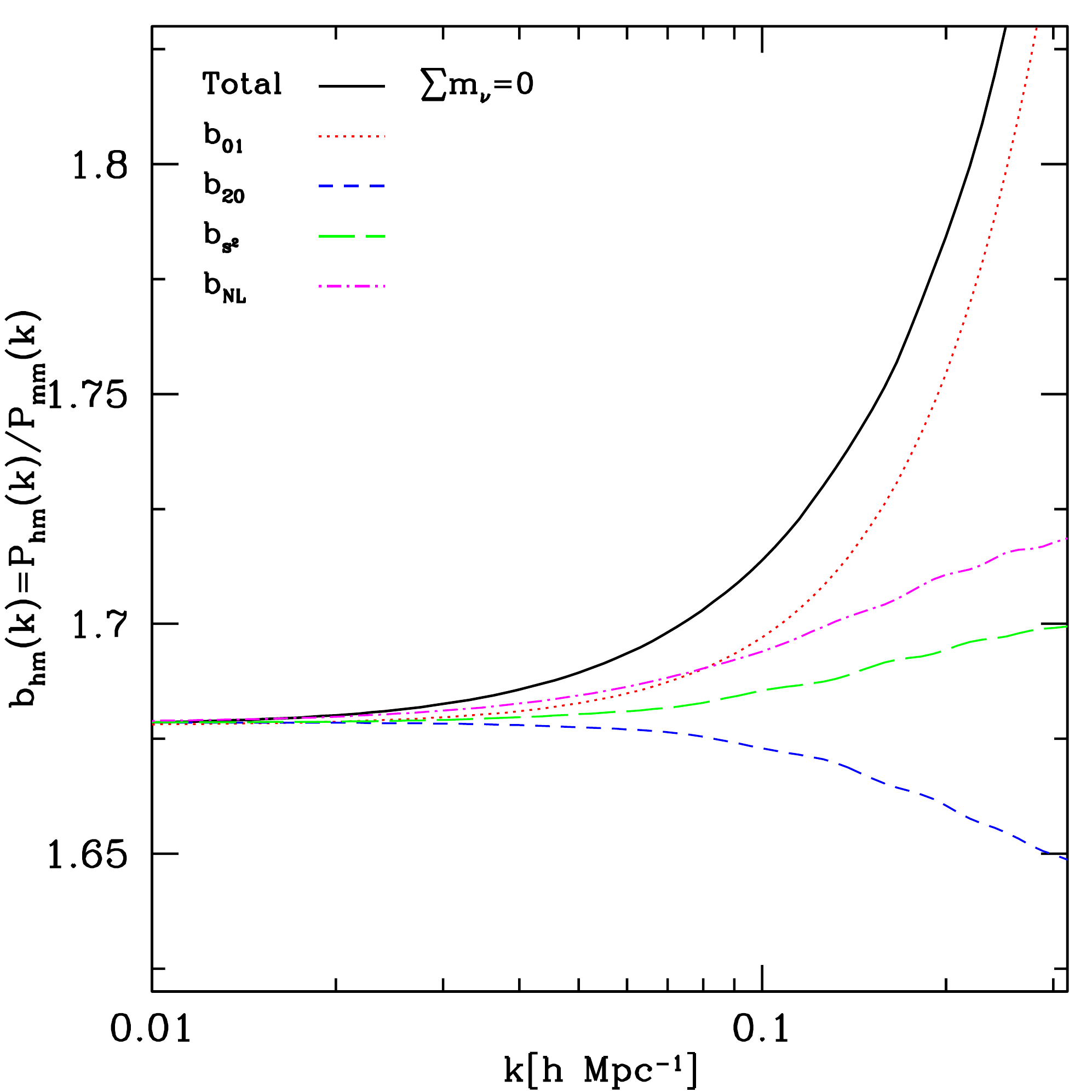}}
\caption{Halo bias at redshift $z=0$ as a function of wavenumber in the case $\sum m_\nu =0$.
The different terms that contribute to the scale-dependence bias in Eq.(\ref{eq:bhm}) are 
labelled according to the bias parameter they are proportional to (see text).
The solid black curve represents the sum of all the contributions, Eq.(\ref{eq:bhm}). All the bias factors have 
been computed consistently from the ESP halo mass function and the relations Eq.(\ref{eq:biask}).}
\label{fig:nonlocal}
\end{figure*}

For this purpose, we compute all the Lagrangian bias factors using the excursion set peak 
(ESP) mass function \cite{Paranjape:2012ks,Paranjape:2012jt}, which has been shown to agree well with simulated 
halo mass functions constructed with a spherical overdensity (SO) criterion 
\cite{Paranjape:2013wna,Biagetti:2013hfa,Hahn:2013gwa}.
Following \cite{ichiki11,Costanzi:2013bha}, we replace the average mass density $\rho_{\rm m}$ by $\rho_{\rm cdm}$, 
and the variance of mass fluctuations $\sigma_{\rm m}$ by $\sigma_{\rm cc}$ in the ESP halo mass function.
While we refer the reader to the aforementioned references for details, it is worth stressing the following 
points:

\begin{enumerate}
\item The bias parameters depend both on redshift and halo mass. 
To follow the analysis of \cite{Villaescusa-Navarro:2013pva,Castorina:2013wga,Costanzi:2013bha} as closely
as possible, we average the Lagrangian bias factors over a suitable range of halo mass,

\be\label{eq:intbias}
b^{\rm L}_{ij} = \frac{ \int_{M_{\rm min}}^{M_{\rm max}} \,b^{\rm L}_{ij}(M)\, n_{\rm ESP}(M)\, {\rm d}M}
{\int_{M_{\rm min}}^{M_{\rm max}}\, n_{\rm ESP}(M)\, {\rm d}M},
\ee
where $n_{\rm ESP}$ is the ESP halo mass function. 
The mass range is $2\times 10^{13} h^{-1} M_{\odot} < M < 3\times 10^{15}h^{-1} M_{\odot}$. In the following Table \ref{tab:sets} some typical values for the bias parameters for different neutrino masses.

\begin{center}\label{tab:sets}
\begin{table}[h]
\begin{tabular}{| c | c | c | c | c |}
\hline
$\sum_i m_{\nu_i} ({\rm eV})$ & $ b^{\rm L}_{\rm 10}$ & $b^{\rm L}_{\rm 01} $ & $R^2_{v} $ & $b^{\rm L}_{\rm 20} $ \\
\hline
0 & 0.68 & 12.32 & 10.41 & $-$0.32\\
\hline
0.1 & 0.73 & 12.43 & 10.38 & $-$0.28\\
\hline
0.2 & 0.80 & 12.55 & 10.33 & $-$0.22\\
\hline
0.3 & 0.87 & 12.68 & 10.29 & $-$0.15\\
\hline
0.6 & 1.15 & 13.13 & 10.17 & 0.24\\
\hline
\end{tabular}
\caption{Lagrangian bias factors as computed from the ESP mass function for the cosmological models
considered here (they are labeled according to the sum of neutrino masses). The bias parameters,
defined relative to the linear density field extrapolated at $z=0$, are weighted over the mass range 
$2\times 10^{13} h^{-1} M_{\odot} < M < 3\times 10^{15}h^{-1} M_{\odot}$.}
\end{table}
\end{center}

\item Here and henceforth, we will use the word ``local'' when we refer to the simplest bias prescription 
without any scale dependence. In this case, we set $b_{ s^2} =  b_{\psi} = b_{st} = b_{01} = 0$.
Conversely, our scale-dependent bias prescription is summarized by Eq.(\ref{eq:biask}) with $R_v$ computed
from Eq.(\ref{eq:rvpeak}).
We only retain the nonlocal peak bias $b_{01}^{\rm L}$ since i) it is the only nonlocal Lagrangian term for which
we have the time evolution and ii) several second-order peak bias factors induce $k^2$-corrections (see e.g. 
\cite{des10,mat}) that are at least partly degenerate with $(b_{01}^{\rm L}-R_v^2)k^2$ over the range 
of wavenumber considered here.

\item The scale dependencies induced by the peak constraint also propagate to $b_{ s^2}$, $b_{\psi}$, $ b_{ st}$ 
etc. for which the Lagrangian to Eulerian mapping can be derived upon analyzing terms like Eq. (\ref{eq:b01}). 
Since $b_{s^2}$ etc. are small however, we do also not expect significant corrections on the scales of interest.

\end{enumerate}

To compare our predictions with the numerical results of 
\cite{Villaescusa-Navarro:2013pva,Castorina:2013wga,Costanzi:2013bha}, 
we use a set of $(\Omega_{\rm m},\Omega_\nu)$ such that the total mass density 
$\Omega_{\rm m} = \Omega_{\rm c} + \Omega_{\nu}$ is held fixed to 0.2708 while the CDM and neutrino 
density are varied. The sum of neutrino masses is $\sum m_\nu=0$, 0.1, 0.2, 0.3 and 0.6 eV, such that 
$\Omega_\nu$ varies between 0 and 0.0131.
We have also adopted $h=0.7$ for the Hubble rate, $n_s=1$ for the scalar 
spectral index, and $A_s=2.43 \times 10^{-9}$ for the amplitude of primordial scalar perturbations.
The normalization amplitude $\sigma_8$ changes in accordance with the sum of neutrino masses
\cite{radiation}.

\begin{figure*}
\centering 
\resizebox{0.49\textwidth}{!}{\includegraphics{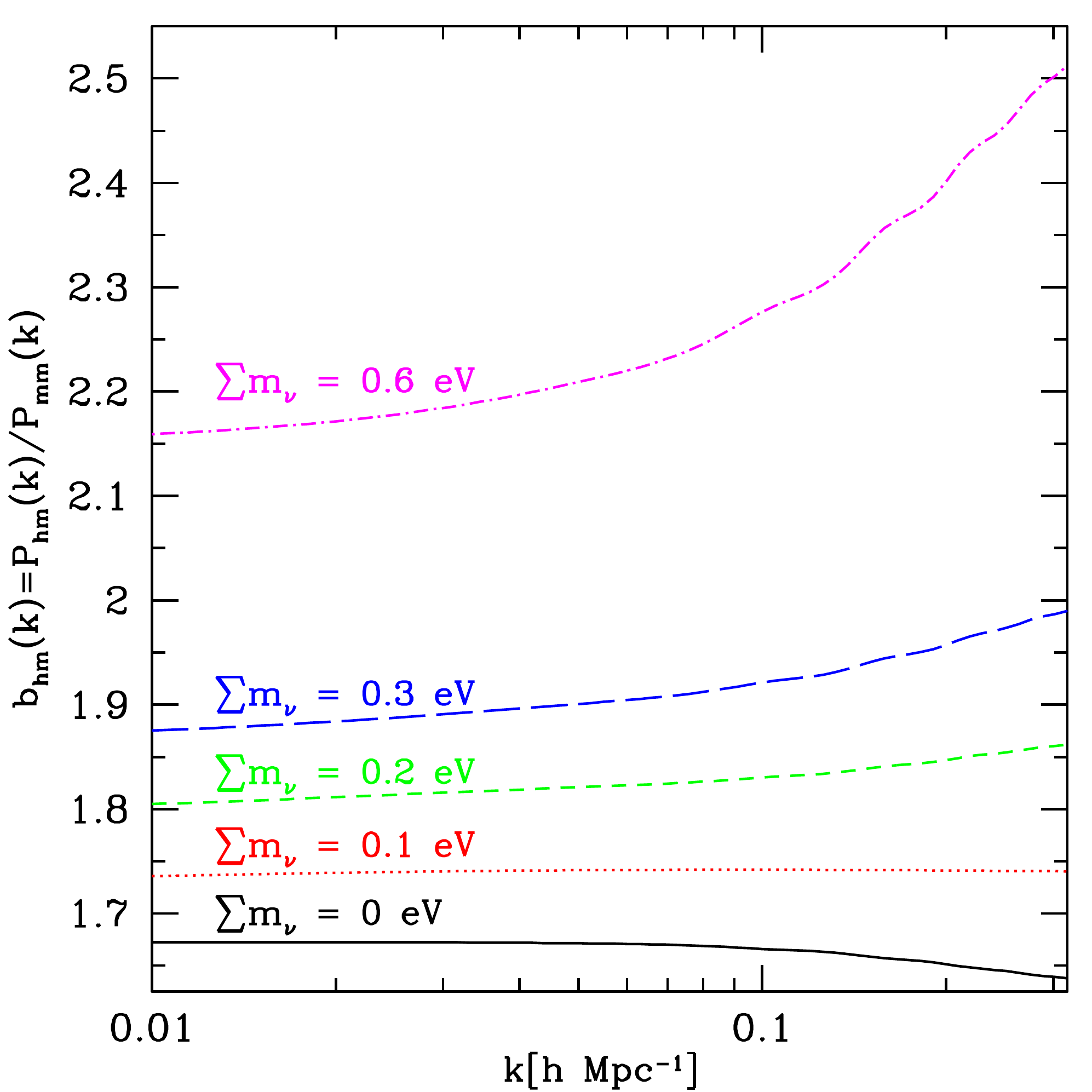}}
\resizebox{0.49\textwidth}{!}{\includegraphics{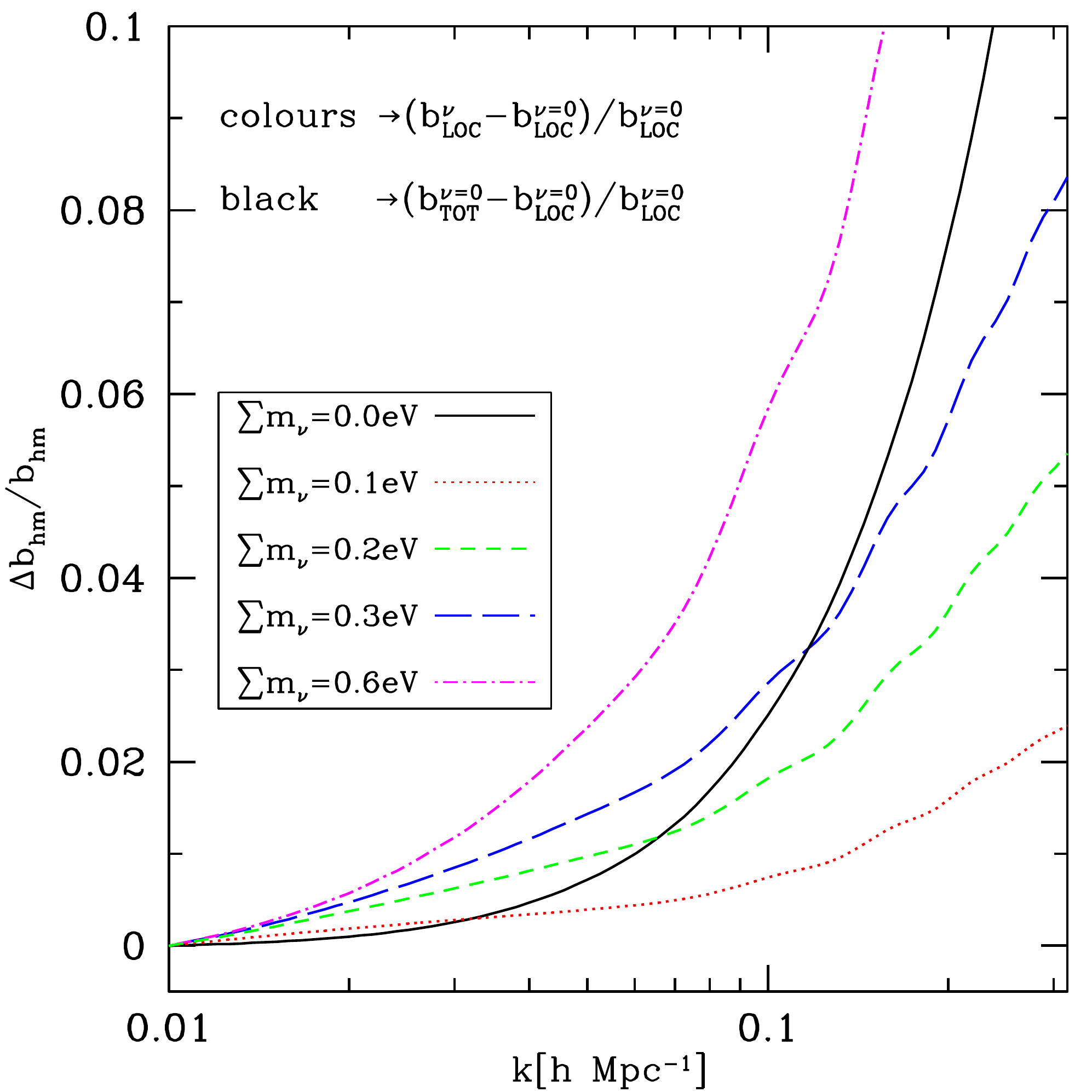}}
\caption{Halo bias (left) and fractional scale-dependence (right) at $z=0$ as a function of wavenumber 
for values of $\sum_i m_{\nu_i}=0$, 0.1, 0.2, 0.3 and 0.6 eV. In the left panel, only the local bias 
terms are included 
in the predictions. In the right panel, the models with non-zero neutrino masses still assume local 
bias, whereas the solid (black) curve represents the case in which massive neutrinos are absent but all
non-local terms are accounted for. The non-local bias contributions induced by gravity and by the peak
constraint generate a sharp rise beyond $k\sim 0.1\, h {\rm  \,Mpc^{-1}}$ substantially steeper than 
the effect of non-zero neutrino mass.}
\label{fig:neutrino}
\end{figure*}

\subsection{Results}
\noindent
The quantity we focus on is the halo bias defined as the ratio of the halo-matter cross-power spectrum to 
the matter auto-power spectrum,

\be\label{eq:bhm}
b_{\rm hm} = \frac{P_{\rm hm}}{P^{\rm NL}_{\rm mm}} =  
\frac{\left(b_{10} + b_{01} k^2\right) P_{\rm cm}^{\rm NL}(k) 
+\Delta P_{\rm hm}(k) + P_{\rm cc}(k) I_3(k)}{P^{\rm NL}_{\rm mm}(k)}.
\ee
It  is expected to be constant (matching the  value of $1+b^{\rm L}_{10}$) on large scales, with a scale 
dependence arising at smaller scales. The origin of this scale-dependence is twofold: the non-locality 
discussed in Sec. \ref{sec:nonlocal} and the suppression of the linear growth rate induced by the presence 
of massive neutrinos. Therefore, we begin by exploring the contributions generated by the non-locality 
arising from gravity and the peak constraint. 
Fig. \ref{fig:nonlocal} displays the various scale-dependent contributions to the halo bias Eq.(\ref{eq:bhm})
when the neutrinos are massless, {\it i.e.} $\sum m_\nu = 0$. We have labelled the curves according to the 
bias parameters that weight each scale-dependent contribution. 
Namely, $b_{20}$ and $b_{s^2}$ denote the two terms of Eq.(\ref{eq:deltaphm}) while $b_{\rm NL}$, defined in Eq. (\ref{bnl}), indicate the term
proportional to $I_3(k)$ in Eq. (\ref{i3}).
Note that $P_{\rm cm}^{\rm NL}(k) = P_{\rm mm}^{\rm NL}(k)$ in this case. Clearly, the dominant contribution 
arises from the peak constraint through the curvature term $(b_{01}^\text{L}-R_v^2)k^2$, which is of positive 
sign for the mass considered here (see Table \ref{tab:sets}).
It is only partly compensated by the other ones, which are negative or close to zero.

\begin{figure*}
\centering 
\resizebox{0.49\textwidth}{!}{\includegraphics{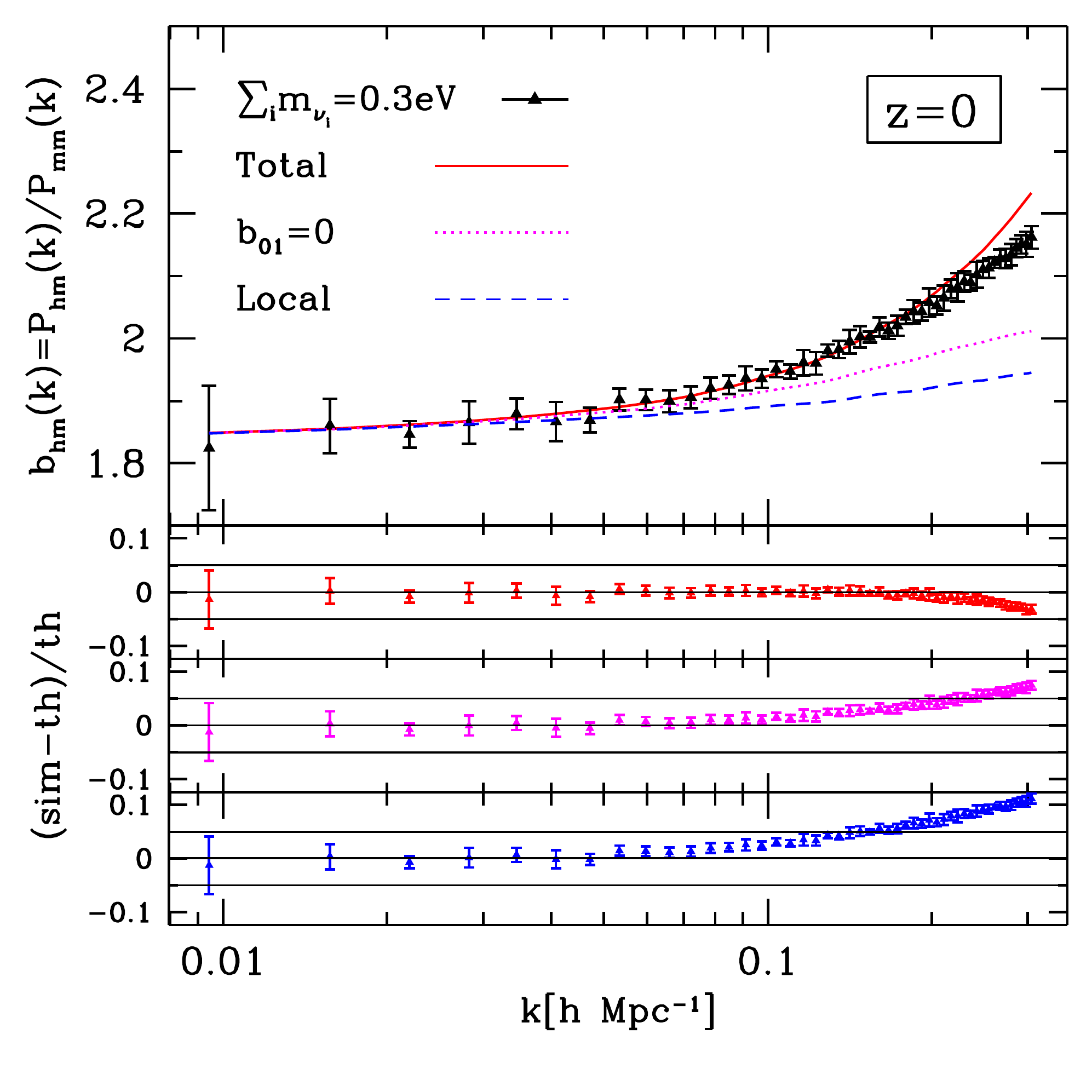}}
\resizebox{0.49\textwidth}{!}{\includegraphics{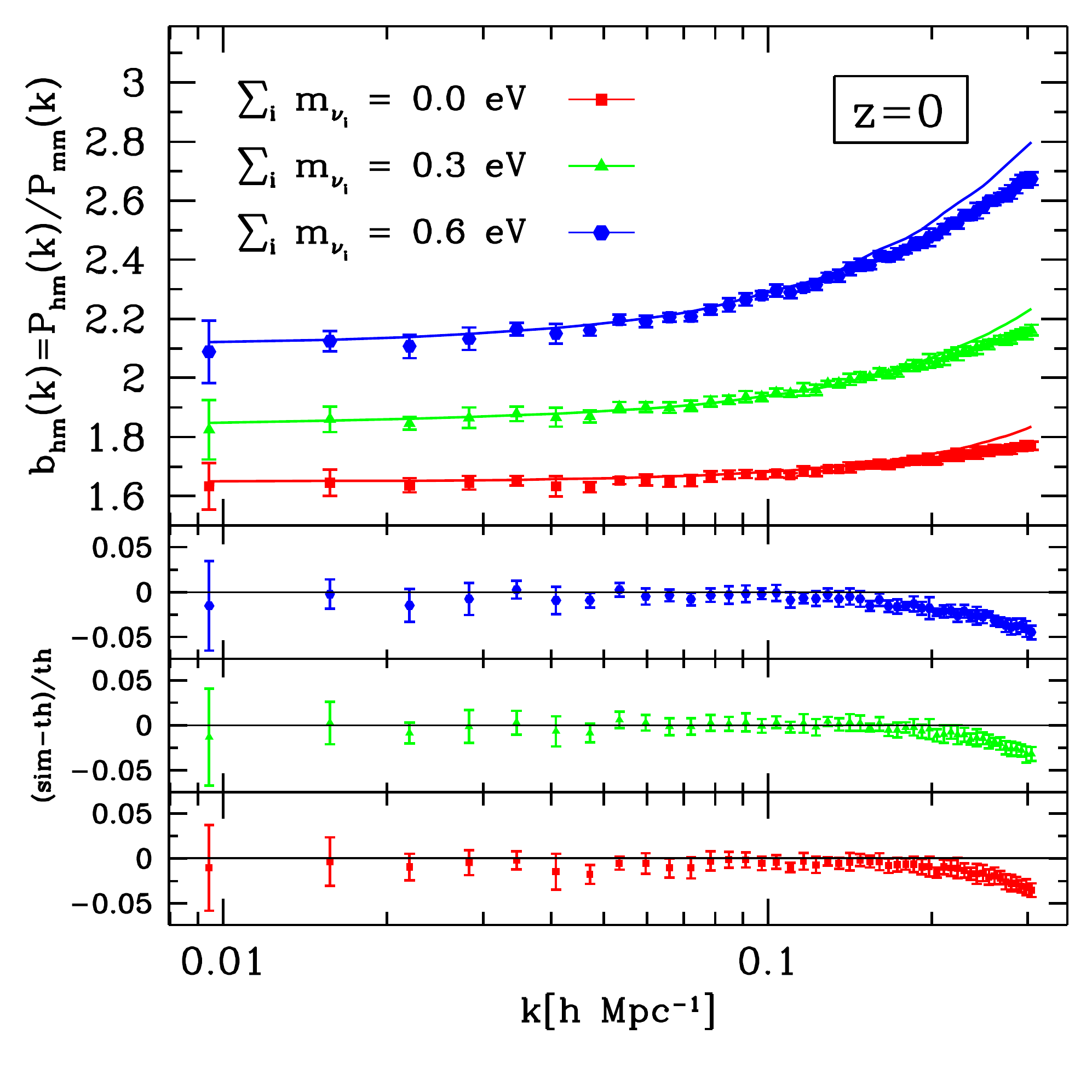}}
\caption{Halo bias at $z=0$ as a function of wavenumber. Data points are from 
\cite{Villaescusa-Navarro:2013pva}. In the upper left panel, we show for $\sum_i m_{\nu_i}=0.3$ eV
the local bias prediction as the dashed (blue) curve, and our full non-local model as the solid (red)
curve. The difference between the solid (red) and the dotted (magenta) curve represents the effect of 
turning off the contribution $b_{01} k^2$ arising from the peak constraint. 
In the upper right panel, we compare our non-local prediction with the numerical data for 
$\sum_i m_{\nu_i}=0$, 0.3 and $0.6$ eV. The lower panels show the fractional deviation between theory 
and simulations. Note that these predictions have no free parameter.}
\label{fig:viel}
\end{figure*}

\begin{figure*}
\centering 
\resizebox{0.49\textwidth}{!}{\includegraphics{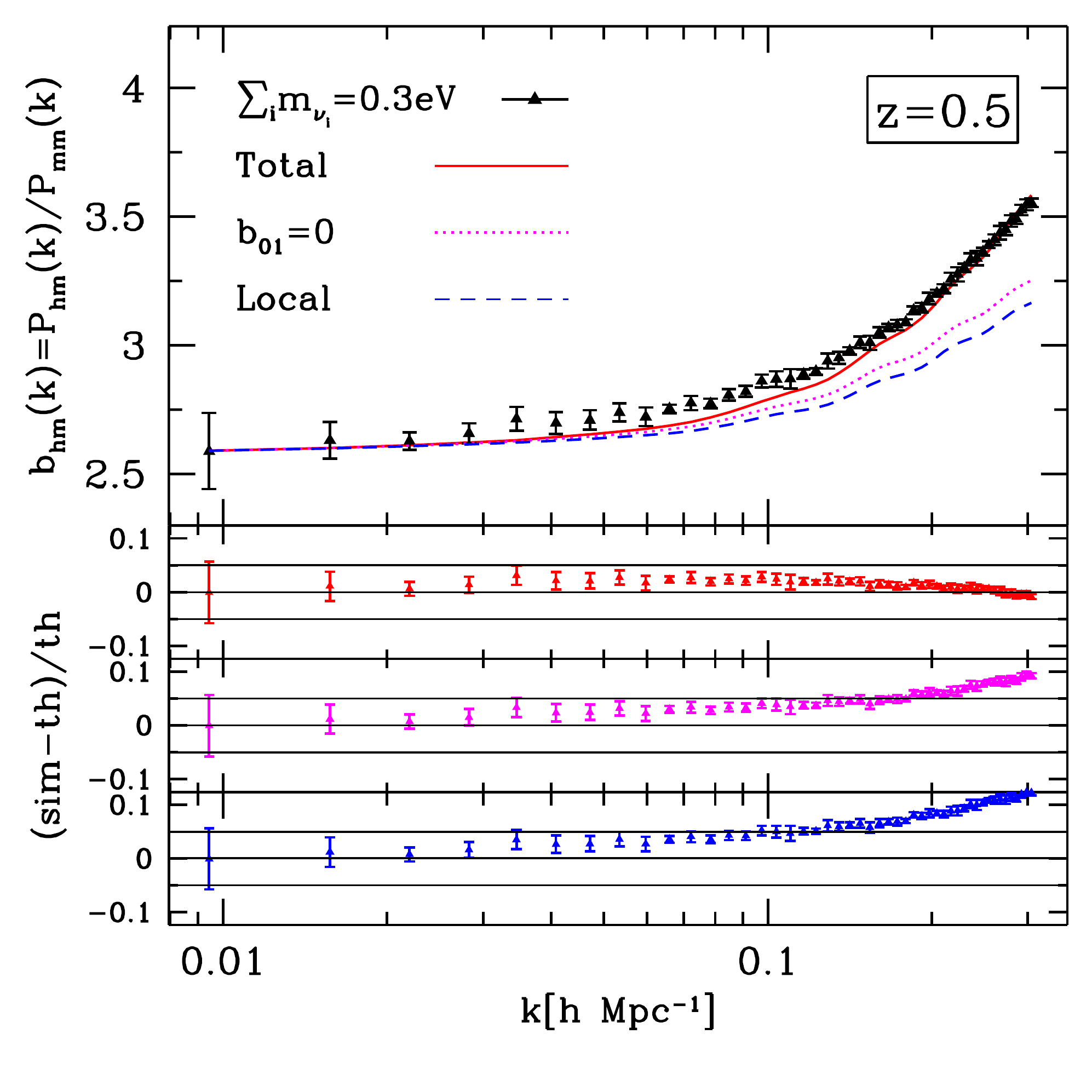}}
\resizebox{0.49\textwidth}{!}{\includegraphics{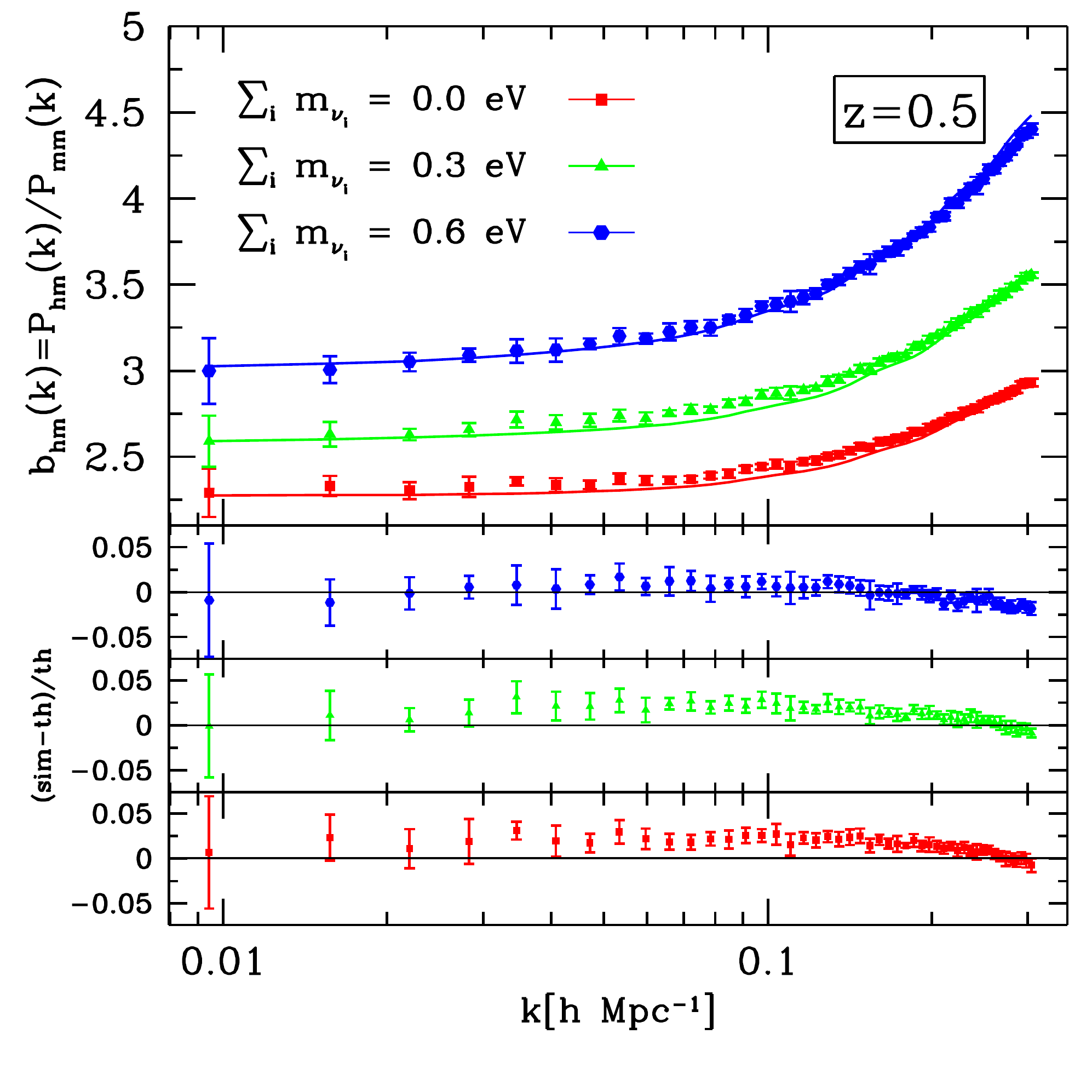}}
\caption{Same as Fig.\ref{fig:viel} but at redshift $z=0.5$. }
\label{fig:viel05}
\end{figure*}
The left panel of Fig. \ref{fig:neutrino} displays the scale-dependence of the halo bias at $z=0$ as a 
function of neutrino masses. For the sake of illustration, we have assumed the simplest local bias model 
specified above, in which only $b_{10}=1+b^{\rm L}_{10}$ and 
$b_{20} = b^{\rm L}_{20} + 8/21\, b^{\rm L}_{10}$ are different from zero. In the right panel of 
Fig.\ref{fig:neutrino}, we show the scale-dependence relative to the $\sum m_\nu=0$ case. This should 
be compared to the solid (black) curve, which represents the scale-dependence obtained when massive 
neutrinos are absent but all the non-local terms are included. These terms generate a sharp rise beyond 
$k\sim 0.1 \, h{\rm \,Mpc^{-1}}$ which is much steeper than the effect of varying the sum of neutrino 
masses. Therefore, a measurement of the scale-dependence of bias over a range of wavenumbers should help
disentangling the contribution of massive neutrinos from that induced by the non-local terms.

In Figs \ref{fig:viel} and \ref{fig:viel05}, we compare our predictions for the halo bias at $z=0$ and 
0.5 with the N-body measurements of \cite{Villaescusa-Navarro:2013pva}. In both figures, the left panel
displays the case of a neutrino sum of 0.3 eV. The dashed (blue) curve is the prediction in the 
simplest local bias model, which falls short of explaining the rise on scales 
$k\gtrsim 0.1 \, h\, {\rm Mpc^{-1}}$. Our full non-local prediction shown as the solid (red) curve agrees
with the numerical data within 3\% for $k\lesssim 0.3 \, h\, {\rm Mpc^{-1}}$. Most of the difference with
the local bias prediction arises from including the peak constraint through the contribution $b_{01} k^2$,
which is the difference between the solid (red) and dotted (magenta) curve. 
The right panel compares our full non-local model with the numerical results for $\sum_i m_{\nu_i}=0$, 
0.3 and $0.6$ eV. In all cases the agreement is at the $\sim 3$\% level down to $k=0.3 \, h\, {\rm Mpc^{-1}}$.
We emphasize again that our theoretical predictions have no free parameter: all the bias factors are 
consistently determined from the ESP halo mass function and from the relations Eq.(\ref{eq:biask}). 
However, the fact that the magnitude of the $k^2$ correction is consistent with $b_{01}^\text{L}-R_v^2$
may be coincidental as we expect similar contributions from higher-order Lagrangian bias. 
Since we do not have as yet Eulerian expressions for these additional bias contributions, we defer a 
thorough discussion of this issue to future work.

\section{Conclusion}\label{sec:conclusion}
In light of the expected accuracy of upcoming galaxy surveys, it is of great importance to characterize as
well as possible the bias between the matter distribution and the luminous tracers. 
The relation between the halo overdensity and the underlying DM fluctuations is not local owing to the 
nonlinear gravitational evolution and to the nature of the initial overdensities which halos collapse form.
In the first part of this paper, we have taken another step towards this goal by computing the non-local 
terms induced by gravity at third-order in perturbation theory. We have computed the coefficients of the 
non-local operators by expanding the halo density contrast in terms of the DM quantities. 
We have generalized our expressions to include a non-local, linear term induced by a peak constraint in the
initial conditions. 
In the second part of the paper, we have applied our results to model the scale-dependence of bias that 
arises in cosmologies with non-zero neutrino masses. 
For the range of halo mass considered here ($M\sim 10^{13} h^{-1} M_{\odot}$), the various non-local bias 
terms conspire to create a steep rise beyond $k\sim 0.1 \, h\, {\rm Mpc^{-1}}$ which is quite distinct from 
the gradual scale-dependence generated by the massive neutrinos. 
We have shown that the inclusion of the non-local bias terms, especially the linear $k^2$ correction induced 
by e.g. a peak constraint in the initial conditions, is crucial for reproducing N-body data if the Eulerian 
bias factors satisfy the relations Eq.(\ref{eq:biask}). Using the ESP halo mass function, we have 
been able to fit the N-body measurement of \cite{Villaescusa-Navarro:2013pva} to within $\sim 3$\% up to 
$k\sim 0.3 \, h\, {\rm Mpc^{-1}}$ without any free parameter. One could envisage further improvements to 
our computation, {\it e.g.} the inclusion of additional Lagrangian bias parameters, to extend the agreement 
to higher wavenumbers. Clearly however, a more detailed comparison including e.g. several halo mass bins, 
bispectrum measurements is in order to test the validity of this approach.

On a final note, we became aware of a similar work by S. Saito et al. \cite{saito14} when completing this 
work. Our findings agree with theirs  wherever there is overlap (basically 
Eq.~(\ref{eq:biasparameters})). In contrast to our approach, Ref.~\cite{saito14} does not include $k^2$-bias.
However, they treat the amplitude of $b_\text{NL}$ as free parameter and include additional constraints from
the bispectrum, so there is no real contradiction. Furthermore, their best-fit values of $b_\text{NL}$ are 
somewhat larger than the expectation $(32/315)\, b_{10}^\text{L}$ for $b_{10}\lesssim 2$, a discrepancy which 
could be partly resolved with the inclusion of $k^2$ bias.

 \section*{Acknowledgments}
\noindent
It is a pleasure to thank M. Viel and F. Villaescusa-Navarro for kindly making their N-body data available
to us, Emiliano Sefusatti for helpful discussions, and S. Saito for correspondence about Ref. \cite{saito14}.
M.B. would also like to thank Enea Di Dio and Francesco Montanari for useful advice about the CLASS code. 
The research of A.K. was implemented under the �Aristeia� Action of the �Operational Programme 
Education and Lifelong Learning� and is co-funded by the European Social Fund (ESF) and National 
Resources. A.K. is also partially supported by European Union�s Seventh Framework Programme 
(FP7/2007-2013) under REA grant agreement n. 329083. A.R. is supported by the Swiss National
Science Foundation (SNSF), project `The non-Gaussian Universe" (project number: 200021140236). 
M.B. and V.D. acknowledge support by the Swiss National Science Foundation.


\begin{small}


\appendix

\section{The halo-mass correlator}\label{ap:hm}

The typical  integral we need to calculate is

\be
{\cal I}_{ij}(k)=\int\frac{\d^3 q}{(2\pi)^3}P_i(q)P_j(| \vec{k}-\vec{q}|)F_2(\vq,\vk-\vq),
\ee
where

\be
F_2(\vq_1,\vq_2)=\frac{5}{7}+\frac{1}{2}\frac{\vq_1\cdot\vq_2}{q_1 q_2}\left(\frac{q_1}{q_2}+\frac{q_2}{q_1}\right)+
\frac{2}{7}\left(\frac{\vq_1\cdot\vq_2}{q_1 q_2}\right)^2.
\ee
Furthermore

\be
S(\vq_1,\vq_2)=\left(\frac{\vq_1\cdot\vq_2}{q_1 q_2}\right)^2-\frac{1}{3}.
\ee
We now define the following variables

\begin{eqnarray}
\mu&=&\frac{\vk\cdot\vq}{kq},\nonumber\\
r&=&\frac{q}{k},
\end{eqnarray}
such that

 {\allowdisplaybreaks
\begin{eqnarray}
\int\frac{\d^3 q}{(2\pi)^3}&=&\frac{k^3}{(2\pi)^2}\int_0^\infty r^2\d r\int_{-1}^{1}\d\mu,\nonumber\\
\eta k&\equiv&| \vec{k}-\vec{q}|=\sqrt{q^2+k^2-2\vk\cdot\vq}=k\sqrt{r^2+1-2r\mu},\nonumber\\
F_2(\vq,\vk-\vq)&=&\frac{5}{7}+\frac{1}{2}\left[(\vk\cdot\vq)-q^2\right]\left(\frac{1}{q^2}+\frac{1}{k^2\eta^2}\right)+\frac{2}{7}\frac{\left[(\vk\cdot\vq)-q^2\right]^2}{q^2 k ^2\eta^2}\nonumber\\
&=&\frac{5}{7}+\frac{1}{2}(\mu r-r^2)\frac{2 r^2+1-2 r \mu}{r^2\eta^2}+\frac{2}{7}\frac{(\mu r-r^2)^2}{r^2 \eta^2}\nonumber\\
&=&\frac{3 r+7\mu-10\mu^2 r}{14 r(1+r^2-2\mu r)}\nonumber\\
S(\vq,\vk-\vq)&=&\frac{(\mu r-r^2)^2}{r^2 \eta^2}-\frac{1}{3}.
\end{eqnarray}
We finally obtain

\be
{\cal I}_{ij}(k)=\frac{k^3}{56\pi^2}\int_0^\infty \d r\int_{-1}^{1}\d\mu\, P_i(k r)P_j\left(k\sqrt{r^2+1-2r\mu}\right)\frac{r(3 r+7\mu-10\mu^2 r)}{(1+r^2-2\mu r)}.
\ee
 Notice  that, in order to avoid the IR divergence when $| \vec{k}-\vec{q}|$ goes to zero, by exploring the symmetry of the integral, we can rewrite it as

\begin{eqnarray}
{\cal I}_{ij}(k)&=&\int\frac{\d^3 q}{(2\pi)^3}P_i(q)P_j(| \vec{k}-\vec{q}|)F_2(\vq,\vk-\vq)
\Theta\left(| \vec{k}-\vec{q}|-q\right)+(i\leftrightarrow j)\nonumber\\
&=&\frac{k^3}{56\pi^2}\int_0^\infty \d r\int_{-1}^{1}\d\mu\, P_i(k r)P_j\left(k\sqrt{r^2+1-2r\mu}\right)\frac{r(3 r+7\mu-10\mu^2 r)}{(1+r^2-2\mu r)}\Theta\left(1-2r\mu\right)+(i\leftrightarrow j),\nonumber\\
&&
\end{eqnarray}
where $\Theta(x)$ is the Heaviside step function.
Similarly

\begin{align}
\int\frac{\d^3 q}{(2\pi)^3}P_i(q)P_j(| \vec{k}-\vec{q}|)F_2(\vq,\vk-\vq)S(\vq,\vk-\vq)&=
\frac{k^3}{168\pi^2}\int_0^\infty \d r\int_{-1}^{1}\d\mu\, P_i(k r)P_j
\left(k\sqrt{r^2+1-2r\mu}\right)
\nonumber \\
&\times \frac{(3 r+7\mu-10\mu^2 r)(1-3\mu^2+4\mu r-2r^2)}{r(1+r^2-2\mu r)^2}.
\end{align}
}

\end{small}


\begin{thebibliography}{99}
\bibitem[\protect\citeauthoryear{}{}]{radiation}{
The simulations were seeded with a transfer function output from a Boltzmann
code at $z=z_i$ rather than at $z=0$, where $z_i$ is the starting redshift, and
subsequently ignore the contribution of radiation to the Hubble rate. Therefore, 
for the sake of comparison we set $\rho_\gamma=0$ in the {\small CLASS} code when 
$0\leq z \leq z_i$. This translates into a few percent increase in the linear 
growth rate at $z=0$, and mainly has an impact on our prediction of $b_{10}$.}
\end{thebibliography}
\end{document}